\documentclass[aps,twocolumn,prb,nopacs,amsmath,amssymb,superscriptaddress]{revtex4-1}
\usepackage[colorlinks,citecolor=blue,linkcolor=cyan]{hyperref}
\usepackage[usenames,dvipsnames,svgnames,table]{xcolor}

\usepackage{dcolumn}
\usepackage{bm}
\usepackage{graphicx}
\usepackage{epstopdf}
\usepackage{color}
\usepackage{placeins}
\usepackage{braket}

\usepackage[usenames,dvipsnames,svgnames]{xcolor}

\begin{document}

\newcommand{\bfk}{{\bf k}}
\newcommand{\bfR}{{\bf R}}
\newcommand{\Hamil}{{\cal H}}

\newcommand{\vc}[1]{\mathbf{#1}}
\newcommand{\e}[1]{\mathrm{e}^{#1}}
\newcommand{\I}{\mathrm{i}}

\title{Topological superconductivity in the extended Kitaev-Heisenberg model}
\author{Johann Schmidt}
 \affiliation{Department of Physics and Astronomy, Uppsala University, Box 516, SE-751 20 Uppsala, Sweden}
\author{Daniel  D.~Scherer}
\affiliation{Niels Bohr Institute, University of Copenhagen, Juliane Maries Vej 30, DK-2100, Denmark}
 \author{Annica M.~Black-Schaffer}
  \affiliation{Department of Physics and Astronomy, Uppsala University, Box 516, SE-751 20 Uppsala, Sweden}
\date{\today}

\begin{abstract}
We study superconducting pairing in the doped Kitaev-Heisenberg model by taking into account the recently proposed symmetric off-diagonal exchange $\Gamma$. By performing a mean-field analysis, we classify all possible superconducting phases in terms of symmetry, explicitly taking into account effects of spin-orbit coupling. Solving the resulting gap equations self-consistently, we map out a phase diagram that involves several topologically nontrivial states. For $\Gamma<0$, we find a competition between a time-reversal symmetry breaking chiral phase with Chern number $\pm1$ and a time-reversal symmetric nematic phase that breaks the rotational symmetry of the lattice. On the other hand, for $\Gamma \geq 0$ we find a time-reversal symmetric phase that preserves all the lattice symmetries, thus yielding clearly distinguishable experimental signatures for all superconducting phases. 
Both of the time-reversal symmetric phases display a transition to a $\mathbb{Z}_2$ non-trivial phase at high doping levels. Finally, we also include a symmetry-allowed spin-orbit coupling kinetic energy and show that it destroys a tentative symmetry protected topological order at lower doping levels. However, it can be used to tune the time-reversal symmetric phases into a $\mathbb{Z}_2$ non-trivial phase even at lower doping.
\end{abstract}
%\pacs{74.20.Mn, 74.20.Rp, 73.20.At, 74.72.-h}
\maketitle

%
% -------------------------------------------------- %
% INTRODUCTION
% -------------------------------------------------- %
% 
\section{Introduction}
The possibility of realizing the Kitaev interaction on the honeycomb lattice with its spin-liquid ground state\cite{Kitaev2006} in a solid state setting\cite{Jackeli2009} has lead to a flurry of experimental and theoretical research.
The necessary combination of lattice geometry, crystal field, spin-orbit coupling, and strong correlations was found to be realized at first in Na$_2$IrO$_3$,\cite{Chaloupka2010} then in Li$_2$IrO$_3$,\cite{Singh2012} and more recently in $\alpha$-RuCl$_3$.\cite{Plumb2014} However, all of these materials were later found to be magnetically ordered at low temperatures,\cite{Singh2010,Sears2015,Williams2016} which highlights the importance of further interactions. In addition to the bond-dependent Kitaev interaction $K$, these materials also exhibit nearest-neighbor Heisenberg exchange $J$\cite{Chaloupka2010} and further, possibly even long-range, interaction terms.\cite{Sizyuk2014,Kimchi2011} Of especial prominence is the symmetric off-diagonal exchange $\Gamma$, which is symmetry allowed in the most general setup.\cite{Rau2014} This term is proven crucial for explaining some of the observed magnetic orderings\cite{Chaloupka2015,Sizyuk2016} and is found to have a significant magnitude in all three compounds.\cite{Winter2016} Taken together, these results point to the importance of studying a $KJ\Gamma$ model, also named the extended Kitaev-Heisenberg model.
Even though the spin-liquid remains elusive in the ground state, there are possible signatures of it above the magnetically ordered phases.\cite{Banerjee2016,Nasu2016} Alternative routes to find spin-liquid phases are therefore currently being explored, such as hydrogen intercalation\cite{Trebst2017} and high-field measurements.\cite{Baek2017, Leahy2017}

Another line of previous works proceeded to look at the superconducting phases produced by the Kitaev interaction on the honeycomb lattice. Since this model is realized in (spin-orbit coupled) Mott insulators, introducing doping is thought to lead to similar physics as in other Mott insulators, such as the doped cuprates.\cite{Lee2006, Edegger2007, Ogata2008, LeHur2009} Initially, two different slave-boson mean-field analyses yielded a time-reversal symmetric\cite{Hyart2012} or a time-reversal breaking phase,\cite{You2012} both with spin-triplet symmetry. Later works were able to reconcile the two, leading to a phase diagram where the time-reversal broken phase appears at very low doping and the time-reversal symmetric one at higher doping in the most common situation of a ferromagnetic Kitaev interaction.\cite{Okamoto2013, Scherer2014} Whereas the time-reversal breaking state always has non-trivial topology with a nonvanishing Chern number,\cite{You2012} the time-reversal symmetric state can be tuned into a topological phase either by doping,\cite{Hyart2012} a Zeeman field,\cite{Hyart2014} or impurities.\cite{Kimme2015} Further including the finite Heisenberg exchange has been shown to lead to a competing superconducting spin-singlet pairing.\cite{Hyart2012} An exotic FFLO pairing state as also been proposed when a specific kind of spin-orbit coupling is dominant.\cite{Liu2016} None of these studies has, however, taken into account the symmetric off-diagonal exchange $\Gamma$, which has been shown to be crucially important for determining the undoped magnetic phase.

In this work we investigate the influence of this symmetric off-diagonal exchange $\Gamma$ on the superconducting phase. We do this by performing a slave-boson mean-field analysis of the extended Kitaev-Heisenberg model.
 We are able to rewrite the off-diagonal exchange in terms of spin-triplet superconducting pairing, which mixes the $\vc{d}$-vector components. A full symmetry analysis of the possible superconducting pairing states, taking into account that the interaction arises from spin-orbit coupled materials, significantly extends the possible odd-parity pairing channels beyond earlier reported analyses.

Performing self-consistent calculations, we find three different superconducting states: i) A chiral solution at intermediate doping for $\Gamma<0$ with a non-zero Chern number. This solution hosts a single chiral Majorana mode on any open boundary. The extension of this region in the phase diagram depends strongly on the size of the interaction parameters. ii) A nematic and time-reversal invariant phase also for $\Gamma<0$ but at higher doping and directly competing with the chiral state. This state breaks the $C_3$ rotation symmetry of the lattice and thus produces an experimental signature that discriminates it sharply from iii), a time-reversal invariant phase for $\Gamma>0$ corresponding to the previously found solution for $\Gamma=0$.\cite{Hyart2012} However, when including a finite $\Gamma$, the $\vc{d}$-vector becomes locked perpendicular to the honeycomb lattice, which breaks the fourfold degeneracy found without $\Gamma$. We find that all these spin-triplet states are stable even when including a small to moderate Heisenberg interaction, that by itself generates spin-singlet superconductivity.

We are also able to topologically classify the time-reversal symmetric states using a $\mathbb{Z}_2$ invariant, which is non-trivial for doping levels above the Lifshitz transition at $\delta =0.25$. In addition, we refine the topological classification by incorporating the appropriate spin-orbit coupling in the kinetic energy. This results in the destruction of a previously discussed symmetry protected topological state found below $\delta=0.25$.\cite{Hyart2014} However, we show that this spin-orbit kinetic term can instead trigger another topological transition into a $\mathbb{Z}_2$ nontrivial state.
Taken together, our results show a remarkable sensitivity of the symmetry and topology of the superconducting phase to the inclusion of the finite $\Gamma$-interaction present in real Kitaev honeycomb materials. These materials thus offer an extraordinary playground for investigating the appearance of, and the transitions between a multitude of different topological superconducting states.

The rest of this paper is organized as follows. We begin by introducing the extended Kitaev-Heisenberg model and its mean-field decoupling in the superconducting channel in Section~\ref{sec:model}. Thereafter, we present a complete symmetry analysis of the spin-triplet superconducting states in Section~\ref{sec:symanal}. In Section~\ref{sec:results} we present our numerical self-consistent calculations. We first focus on the time-reversal breaking state, followed by discussing the different time-reversal symmetric solutions grouped by the parameter regime in which they appear. Finally, we study the influence of the spin-orbit coupled kinetic energy term in Section~\ref{sec:hopping}, before ending with some concluding remarks in Section~\ref{sec:conclusion}.

%
% -------------------------------------------------- %
% Method
% -------------------------------------------------- %
\section{Model and Method \label{sec:model}}

\begin{figure}[tb]
\includegraphics[width=.5\textwidth]{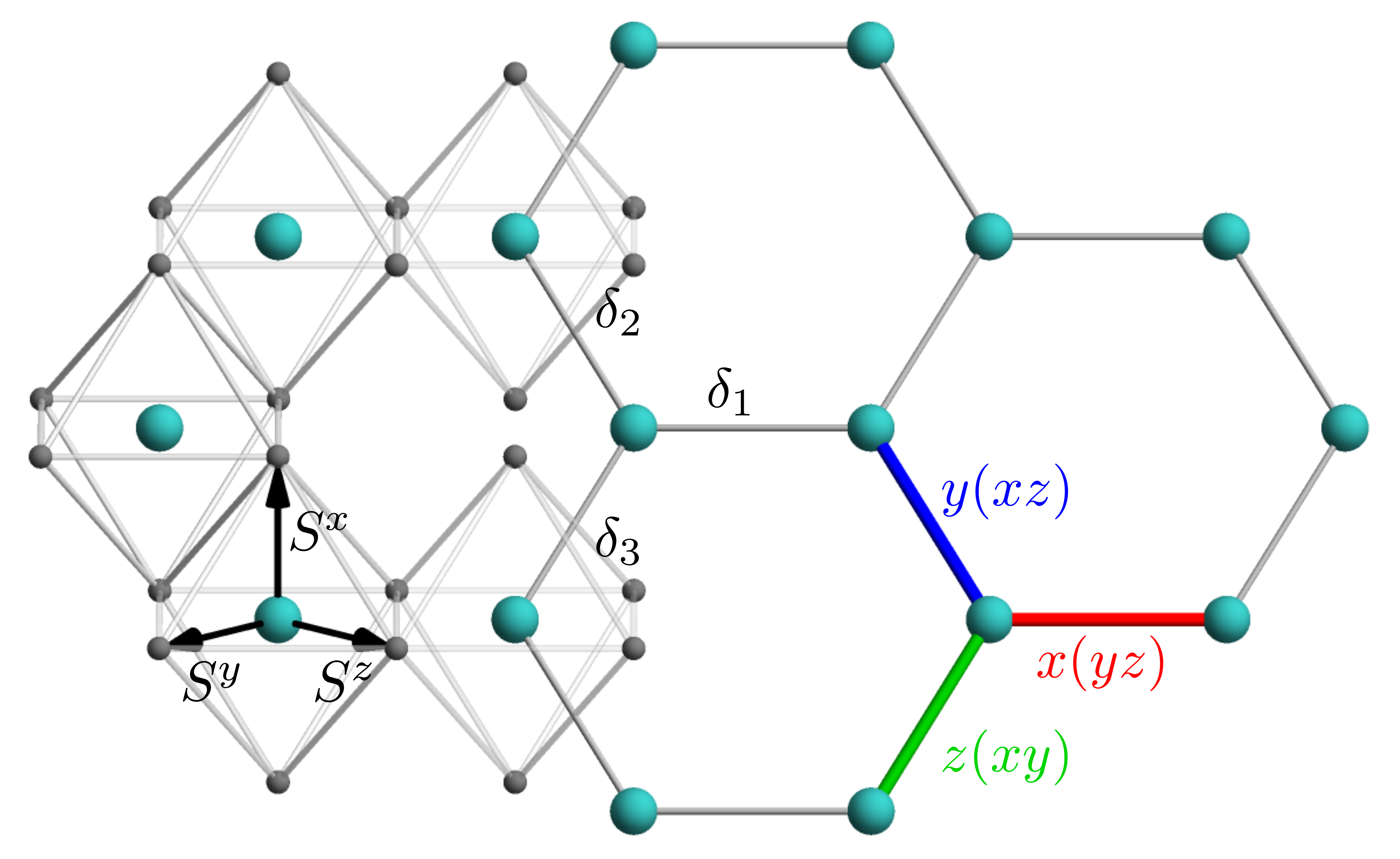}
\caption{Sketch of the honeycomb lattice with the octahedral cages typical for materials hosting the extended Kitaev-Heisenberg model. The Kitaev interaction on the red, blue, and green bonds involve the x, y, and z-component, respectively, while the other two spin components in parenthesis are active in the off-diagonal exchange on the corresponding bond. Black arrows mark the spin coordinate system. The spin component involved in the Kitaev exchange on a particular bond is always perpendicular to the corresponding nearest neighborg vector $\delta_i$.\label{fig:bonds}}
\end{figure}

Considering extended Kitaev-Heisenberg materials as Mott insulators at half filling, we aim to study the possible superconducting instabilities arising upon the introduction of additional charge carriers. To this end we consider the Hamiltonian
\begin{align}
\label{eq:Htot}
H &= H_{\mathrm{k}} + H_{\mathrm{KJ\Gamma}}
\end{align}
on the honeycomb lattice, with a kinetic term $H_\mathrm{k}$ that encapsulates the doping and the interaction Hamiltonian $H_{\mathrm{KJ\Gamma}}$, which consists of three terms:\cite{Rau2014} the Kitaev term $K$, which is a bond-dependent Ising-like interaction, an isotropic Heisenberg interaction between nearest-neighbor spins of strength $J$, and finally a symmetric off-diagonal interaction of strength $\Gamma$, which couples the spin components that are not involved in the Kitaev interaction. It can be written as
\begin{align}
\label{eq:extKHHam}
H_{\mathrm{KJ\Gamma}} &= J \sum_{\braket{i,j}} \left( \vc{S}_i \cdot \vc{S}_j - \frac{1}{4} n_i n_j \right)\nonumber\\
&+ K \sum_{\braket{i,j}} S_i^{\gamma(i,j)}S_j^{\gamma(i,j)}\\
&+ \Gamma \sum_{\braket{i,j}} \left( S_i^{\alpha(i,j)}S_j^{\beta(i,j)} + S_i^{\beta(i,j)}S_j^{\alpha(i,j)}\right),\nonumber
\end{align}
where $\vc{S}_i$ represent the effective spin moment $j_{\rm eff} = \frac{1}{2}$ present at every site $i$ of the honeycomb lattice. Furthermore, $\alpha(i,j)$, $\beta(i,j)$, and $\gamma(i,j) = x,y,z$, depending on the nearest-neighbor bond between sites $i$ and $j$ (see Fig.~\ref{fig:bonds} for details), where $\alpha(i,j) \neq \beta(i,j) \neq \gamma(i,j)$, while $n_i=c^\dagger_{i,\sigma,o}c_{i,\sigma,o}$ is the electron density operator. All sums run only over the three nearest-neighbor bonds $\boldsymbol{\delta}_1 = \left(1,0\right),\boldsymbol{\delta}_2= \frac{1}{2}\left(-1, \sqrt{3}\right),$ and $\boldsymbol{\delta}_3= \frac{1}{2}\left(-1,-\sqrt{3}\right)$, in units of the nearest-neighbor distance.

It is possible to recast the extended Kitaev-Heisenberg Hamiltonian in a form more convenient for the study of superconducting pairing, by introducing spin-singlet and -triplet operators defined on the nearest-neighbor bonds $\braket{i,j}$
\begin{align}
s^\dagger_{ij} &=\frac{1}{\sqrt{2}} \sum_{\sigma, \bar{\sigma}} c_{i,\sigma,a}^\dagger c_{j,\bar{\sigma},b}^\dagger \I \left( \mathbf{\sigma_y} \mathbf{\sigma_0} \right)_{\sigma,\bar{\sigma}},\nonumber\\
t^{\alpha^\dagger}_{ij} &=\frac{1}{\sqrt{2}} \sum_{\sigma, \bar{\sigma}} c_{i,\sigma,a}^\dagger c_{j,\bar{\sigma},b}^\dagger \I \left( \mathbf{\sigma_y} \mathbf{\sigma_\alpha} \right)_{\sigma,\bar{\sigma}},
\end{align}
where $\alpha=x,y,z$, and $\sigma$ are the Pauli matrices acting on spin space, with $\sigma_0$ being the $2\times 2$ identity matrix.

Because of the density-density term, introduced to include doping effects, the Heisenberg term can then be written purely in terms of the singlet operators\,\cite{Baskaran2002, BlackSchaffer2007}
\begin{align*}
\vc{S}_i \cdot \vc{S}_j - \frac{1}{4} n_i n_j = - s^\dagger_{ij} s_{ij}.
\end{align*}
Further, the Kitaev term on the $z$-bond takes the form\cite{Hyart2012}
\begin{align*}
S_i^z S_j^z = \frac{1}{4} \left( -s_{ij}^\dagger s_{ij} + t_{ij}^{x^\dagger} t_{ij}^x + t_{ij}^{y^\dagger} t_{ij}^y -t_{ij}^{z^\dagger} t_{ij}^z\right)
\end{align*}
(for the $x$ and $y$-bond, the negative sign appears in front of the respective triplet term). Finally, we also rewrite the off-diagonal terms with the help of the triplet operators:
\begin{align*}
& S_i^{\alpha}S_j^{\beta} + S_i^{\beta}S_j^{\alpha}
= \frac{1}{2} \left( t^{\alpha^\dagger}_{ij} t^{\beta}_{ij} + t^{\beta^\dagger}_{ij} t^{\alpha}_{ij} \right).
\end{align*}

In a next step we perform a mean-field decoupling by replacing the singlet and triplet operators by their expectation values in the usual way treating superconductivity. For complete generality, we retain independent order parameters on each bond, such that there are in total 12 different order parameters. This allows us to fully capture the orbital dependence of the superconducting order. Three of these 12 are singlet order parameters, one for each nearest-neighbor bond, which we combine in a vector $\vc{\Delta} = (\Delta_1,\Delta_2,\Delta_3)$. The remaining nine are triplet order parameters, which make up the usual $\vc{d}$-vector, with each component $\vc{d}^\alpha$ containing three nearest-neighbor bond order parameters $\vc{d^\alpha} = (d^\alpha_1, d^\alpha_2, d^\alpha_3)$. For concise notation, we compile all nine triplet order parameters in a matrix of the form
\begin{align}
\label{eq:matrix}
\vc{d} = \left(
\begin{matrix}
d^x_1 & d^x_2 & d^x_3\\
d^y_1 & d^y_2 & d^y_3\\
d^z_1 & d^z_2 & d^z_3
\end{matrix}
\right),
\end{align}
with each row representing one $\vc{d}$-vector component and each column one of the three nearest-neighbor bonds. The resulting mean-field Hamiltonian is given by
\begin{align}
\label{eq:HDelta}
H_{\mathrm{\Delta}} &= \sum_{\braket{i,j}} \left( \Delta_{ij} s^\dagger_{ij} + \sum_{\alpha} d^\alpha_{ij} t^{\alpha\dagger}_{ij} +\mathrm{h.c.} \right),
\end{align}
where we have dropped constant terms that only change the overall energy of the system. The order parameters in Eq.~\eqref{eq:HDelta} are defined via the self-consistency equations
\begin{widetext}
\begin{align}
\vc{\Delta} &= \frac{1}{\sqrt{2}}\left( -J-\frac{K}{4} \right) \left(\braket{s_{i \delta_1}},
\braket{s_{i \delta_2}},
\braket{s_{i \delta_3}}
\right),
\nonumber\\
\label{eq:selfcons}
 \mathbf{d^x} &= \frac{1}{\sqrt{2}}\left(
- \frac{K}{4} \braket{t_{i \delta_1}^x},
\frac{K}{4} \braket{t_{i \delta_2}^{x}} + \frac{\Gamma}{2} \braket{t_{i \delta_2}^{z}},
\frac{K}{4} \braket{t_{i \delta_3}^{x}} + \frac{\Gamma}{2} \braket{t_{i \delta_3}^{y}}
\right),\\
\mathbf{d^y} &= \frac{1}{\sqrt{2}}\left(
\frac{K}{4} \braket{t_{i \delta_1}^{y}} + \frac{\Gamma}{2} \braket{t_{i \delta_1}^{z}},
- \frac{K}{4} \braket{t_{i \delta_2}^y},
\frac{K}{4} \braket{t_{i \delta_3}^{y}} + \frac{\Gamma}{2} \braket{t_{i \delta_3}^{x}}
\right),
\nonumber\\
\mathbf{d^z} &= \frac{1}{\sqrt{2}}\left(
\frac{K}{4} \braket{t_{i \delta_1}^{z}} + \frac{\Gamma}{2} \braket{t_{i \delta_1}^{y}},
\frac{K}{4} \braket{t_{i \delta_2}^{z}} + \frac{\Gamma}{2} \braket{t_{i \delta_2}^{x}},
-\frac{K}{4} \braket{t_{i \delta_3}^z}
\right).\nonumber
\end{align}
\end{widetext}
Here the vectors run over the three nearest-neighbor bonds. We see directly that the Heisenberg term $J$ only gives rise to singlet pairing, whereas the Kitaev $K$ and off-diagonal $\Gamma$ exchange terms generate triplet pairing. Moreover, $\Gamma$ couples two different triplet components on the same bond, e.g.~the $x$ and $z$-triplet on the $\delta_2$ bond. Including only $K$ does not generate such coupling. Overall, this leads to a competition between triplet states driven by the $K$ and $\Gamma$ terms and singlet states from the $J$ interaction. We also expect from Eq.~\eqref{eq:selfcons}, that including $\Gamma$ will lead to a much more complex triplet state.

At half-filling, the extended Kitaev-Heisenberg interaction is the effective interaction in a (spin-orbit) Mott insulator. It is thus reasonable to expect finite doping to induce superconductivity, similar to the situation considered in other Mott insulators.\cite{Lee2006,Edegger2007,Ogata2008,LeHur2009} Taking this starting point, we model the kinetic part of the Hamiltonian by the tight-binding Hamiltonian
\begin{align}
\label{eq:Hk}
H_{\mathrm{k}} &= -\tilde{t} \sum_{\braket{i,j},\sigma} \left( c^\dagger_{i,\sigma,a} c_{j,\sigma,b} + h.c. \right) + \tilde{\mu} \sum_{i,\sigma,o} \left( c^\dagger_{i,\sigma,o} c_{i,\sigma,o}\right),
\end{align}
where $c_{i,\sigma,o}^\dagger$ creates an electron at site $i$ on sublattice $o=a,b$ with (pseudo-)spin $\sigma$. The hopping $\tilde{t}$ is restricted to nearest-neighbor bonds $\braket{i,j}$, which gives rise to the well-known graphene-like band structure with Dirac cones at the Brillouin zone points $K$ and $K'$. To exclude the double site-occupancy, we include the Gutzwiller approximation in $\tilde{t}$ through a $U(1)$ slave-boson mean-field approach, which leads to a rescaling of the effective hopping amplitude.\cite{Barnes1976, Baskaran1987,Lee2006,Edegger2007,Ogata2008,LeHur2009} Thus, $\tilde{t} = t \delta$, where $t$ is the bare hopping parameter, and $\delta$ corresponds to the hole doping level, such that the number of electrons per site is given by $1-\delta$. This approximation also requires an adjustment of the chemical potential $\tilde{\mu}$ for each $\delta$, which we perform by calculating the filling at $\tilde{\mu}$ and demanding it to equal to $1-\delta$. The same approach has previously been successfully applied both for the Heisenberg\cite{BlackSchaffer2007, Honerkamp2008, Wu2013} and Kitaev-Heisenberg\cite{Hyart2012,Scherer2014} interactions on the honeycomb lattice.
We finally comment that the kinetic energy term in Eq.~\eqref{eq:Hk} does not explicitly take into account any spin-orbit effects on the kinetic energy. The strong spin-orbit coupling present in materials described by the extended Kitaev-Heisenberg Hamiltonian is of course included in-so-far as it leads to the formation of the effective $j_\mathrm{eff}=\frac{1}{2}$ states and their anisotropic $K$ and $\Gamma$ interactions. In Section~\ref{sec:hopping} we also consider how including a symmetry-allowed spin-orbit driven hopping influences the superconducting states.

%
% -------------------------------------------------- %
% Results
% -------------------------------------------------- %
\section{Symmetry analysis \label{sec:symanal}}
We start the analysis of superconductivity in the extended Kitaev-Heisenberg model by performing a classification of the superconducting order parameters. This can always be done in terms of the irreducible representations (irreps) of the point group of the model, as they determine the possible solutions of the self-consistent non-linear gap equation.

The kinetic part of the Hamiltonian is symmetric under the symmetries of the point group $D_{6h}$, but the symmetry of the interaction, and subsequently also the mean-field Hamiltonian, is reduced to $D_{3d}$, due to the bond dependence of the interaction. The classification furthermore needs to take into account, that the materials realizing the extended Kitaev-Heisenberg model exhibit strong spin-orbit coupling, which is implicitly taken into account from the very start when writing down the interaction Hamiltonian Eq.~\eqref{eq:extKHHam}. This requires a locking of the orbital and spin symmetry transformations together during the classification.\cite{Sigrist1991} For example, a $C_3$ rotation of the lattice has to be performed together with an equivalent rotation around the $(1,1,1)$ axis in spin space to leave the Hamiltonian Eq.~\eqref{eq:Htot} invariant. With the choice of spin coordinate system as presented in Fig.~\ref{fig:bonds}, the $(1,1,1)$ axis namely points perpendicular to the honeycomb plane, and a $C_3$ rotation around this axis therefore maps $S^x \rightarrow S^y \rightarrow S^z \rightarrow S^x$. As a consequence, this spin-orbit coupling leads to a mixing of the three triplet components in the basis functions of the irreps.

For the purpose of studying the effect of the off-diagonal exchange, only the triplet order parameters need to be classified. The analysis of the spin-singlet pairing is not affected by the spin-orbit coupling and has been discussed earlier.\cite{BlackSchaffer2007}
For the nine triplet order parameters, there are in total nine basis functions: One for the $A_{1u}$ irrep, two transforming according to $A_{2u}$, and six that correspond to $E_u$. Using the matrix notation introduced in Eq.~\eqref{eq:matrix}, the nine basis functions are:
\begin{widetext}
\begin{align}
\vc{d}_{\mathrm{A_{1u}}} &= \left(
\begin{matrix}
0 & -1 & 1\\
1 & 0 & -1\\
-1 & 1 & 0
\end{matrix}
\right),
&
\vc{d}_{\mathrm{A_{2u},1}} &= \left(
\begin{matrix}
1 & 0 & 0\\
0 & 1 & 0\\
0 & 0 & 1
\end{matrix}
\right),
&
\vc{d}_{\mathrm{A_{2u},2}} &= \left(
\begin{matrix}
0 & 1 & 1\\
1 & 0 & 1\\
1 & 1 & 0
\end{matrix}
\right),
\nonumber
\\
\label{eq:symclass}
\vc{d}_{\mathrm{E_{u},1}} &= \left(
\begin{matrix}
1 & 0 & 0 \\
0 & 0 & 0 \\
0 & 0 & -1
\end{matrix}
\right),
&
\vc{d}_{\mathrm{E_{u},2}} &= \left(
\begin{matrix}
1 & 0 & 0 \\
0 & -1 & 0 \\
0 & 0 & 0
\end{matrix}
\right),
&
\vc{d}_{\mathrm{E_{u},3}} &= \left(
\begin{matrix}
0 & 1 & 0 \\
0 & 0 & -1 \\
0 & 0 & 0
\end{matrix}
\right),
\\
\vc{d}_{\mathrm{E_{u},4}} &= \left(
\begin{matrix}
0 & 1 & 0 \\
0 & 0 & 0 \\
-1 & 0 & 0
\end{matrix}
\right),
&
\vc{d}_{\mathrm{E_{u},5}} &= \left(
\begin{matrix}
0 & 0 & 1 \\
0 & 0 & 0 \\
0 & -1 & 0
\end{matrix}
\right),
&
\vc{d}_{\mathrm{E_{u},6}} &= \left(
\begin{matrix}
0 & 0 & 1 \\
-1 & 0 & 0 \\
0 & 0 & 0
\end{matrix}
\right).
\nonumber
\end{align}
\end{widetext}
Here we have chosen to present them in a non-normalized and non-orthogonal representation, which simplifies the identification of the solutions obtained from the self-consistent calculations.

The six $E_u$ basis functions can be identified as being constructed from three different sets ($\{1,2\},\{3,4\},\{5,6\}$) that are not related by symmetry, and which do not correspond directly to the $\vc{d}$-vector components. All basis functions of this irrep will be degenerate and can generally form some linear combination at $T_c$. An intuitive picture of the different sets can be gained by considering the diagonal and the two off-diagonals of the matrix presentation
\begin{align*}
\left(\begin{matrix}
d^x_1 & 0 & 0\\
0 & d^y_2 &0\\
0 & 0 & d^z_3
\end{matrix}\right), &&
\left(\begin{matrix}
0 & d^x_2 & 0\\
0 & 0 & d^y_3\\
d^z_1 & 0 & 0
\end{matrix}\right) &&
\left(\begin{matrix}
0 & 0 & d^x_3\\
d^y_1 & 0 & 0\\
0 & d^z_2 & 0
\end{matrix}\right).
\end{align*}
Each of these (off-)diagonals is linked by the $C_3$ rotation that simultaneously rotates the $\vc{d}$-vector in spin space and the three nearest-neighbor bonds. The resulting mapping consists of $d^x \rightarrow d^y \rightarrow d^z \rightarrow d^x$ and $\vc{\delta_1} \rightarrow \vc{\delta_2} \rightarrow \vc{\delta_3} \rightarrow \vc{\delta_1}$ and thus does not relate these (off-)diagonals to each other. As a consequence, the symmetry analysis has to necessarily treat the full $\vc{d}$-vector instead of its individual components.
Using these (off-)diagonals we see how the $A_{2u}$ irreps are fully symmetric, i.e.~its components are always 1, whereas the $A_{1u}$ irrep adds an additional minus sign between different off-diagonals.
In the $E_u$ irrep the (off-)diagonals follow a $(1,0,-1)$ pattern that makes them odd under the application of the $C_3$ rotation. This pattern on nearest-neighbor bonds is actually equivalent to that of the singlet order parameters belonging to the $E_{2g}$ irrep of the $D_{6h}$ point group.\cite{BlackSchaffer2007}

We can in fact already make some qualitative statements about the expected symmetries by using the basis functions as an ``one-shot'' input into the self-consistency equations, Eqs.~\eqref{eq:selfcons}. The basis function $\vc{d}_{A_{1u}}$ is an eigenfunction of this operation, with eigenvalue $\frac{K}{4}-\frac{\Gamma}{2}$, so we can likely expect it to be stable for $K<0$ and $\Gamma>0$. The diagonal function $\vc{d}_{A_2u,1}$, involving only diagonal entries, is also an eigenfunction, but with eigenvalue $-\frac{K}{4}$, such that it should be stable for $K>0$. Other eigenvectors can be constructed from the difference between two basis functions of the $E_u$ irrep. The linear combination $\vc{d}_{E_{u,3}}-\vc{d}_{E_{u,5}}$, for example, is an eigenvector with eigenvalue $\frac{K}{4}+\frac{\Gamma}{2}$, which indicates that this linear combination might be favored at $K<0$ and $\Gamma<0$. Since several of the $E_u$ basis functions are degenerate at $T_c$, the exact solution is likely some kind of of linear combination of them, but the exact combination can only  be determined by fully solving the self-consistent equations numerically. Still, we can from these simple arguments already predict that the symmetry of the superconducting state is very strongly influenced by the sign of $\Gamma$.

\section{Self-consistent Phase Diagram \label{sec:results}}
Having performed the general symmetry analysis in the previous section, we now turn to actually solving for the superconducting order parameters. We do this by diagonalizing the mean-field Hamiltonian
\begin{align*}
H = H_\mathrm{k} + H_\mathrm{\Delta}
\end{align*}
using a random starting order parameter and then calculating the expectation values in the self-consistency conditions Eqs.~\eqref{eq:selfcons}. The thus obtained new order parameters are fed back to the Hamiltonian and we iterate the procedure until we reach full self-consistency.

The exact values of the extended Kitaev-Heisenberg model for either of the materials, which are discussed to realize it, are still under debate.\cite{Yamaji2014, Chaloupka2015, Hu2015, Kim2015-RuCl, Banerjee2016,  Yamaji2016, Winter2016, Nishimoto2016, Ran2017, Janssen2017} Furthermore, several studies have shown that the parameters are very sensitive to deformations of the lattice.\cite{Sizyuk2014, Katukuri2014, Yadav2016, Kim2016}
Only some very general conclusions can be drawn at the current moment: The Kitaev interaction $K$ is usually dominant and ferromagnetic, the Heisenberg interaction $J$ is subdominant and mainly antiferromagnetic, and the off-diagonal exchange $\Gamma$ can have either sign, but is usually smaller than $K$.
We thus restrict our parameter space by setting $K = -t$ and varying the strength of the off-diagonal interaction between $-t < \Gamma < t$. In a second step, we study the stability of the observed solutions when switching on a finite Heisenberg interaction with $0< J<t$, which generates a competition between spin-triplet an singlet superconductivity. This choice of parameters also allows us to connect to previous results obtained at $\Gamma=0$.\cite{Hyart2012} The hole doping is adjusted by varying the doping level $\delta$ in the interval between $0.1$ and $0.3$, which includes a van Hove singularity in the density of states at $\delta = 0.25$. It has been shown in previous works, that the decoupling scheme used here produces reliable result at such not too small doping levels.\cite{Okamoto2013, Scherer2014} At very low values of $\delta$, the ratio between interaction terms and the hopping can become artificially large within our scheme, possibly resulting in numerical instability.

\begin{figure}[tb]
\includegraphics[width=.4\textwidth]{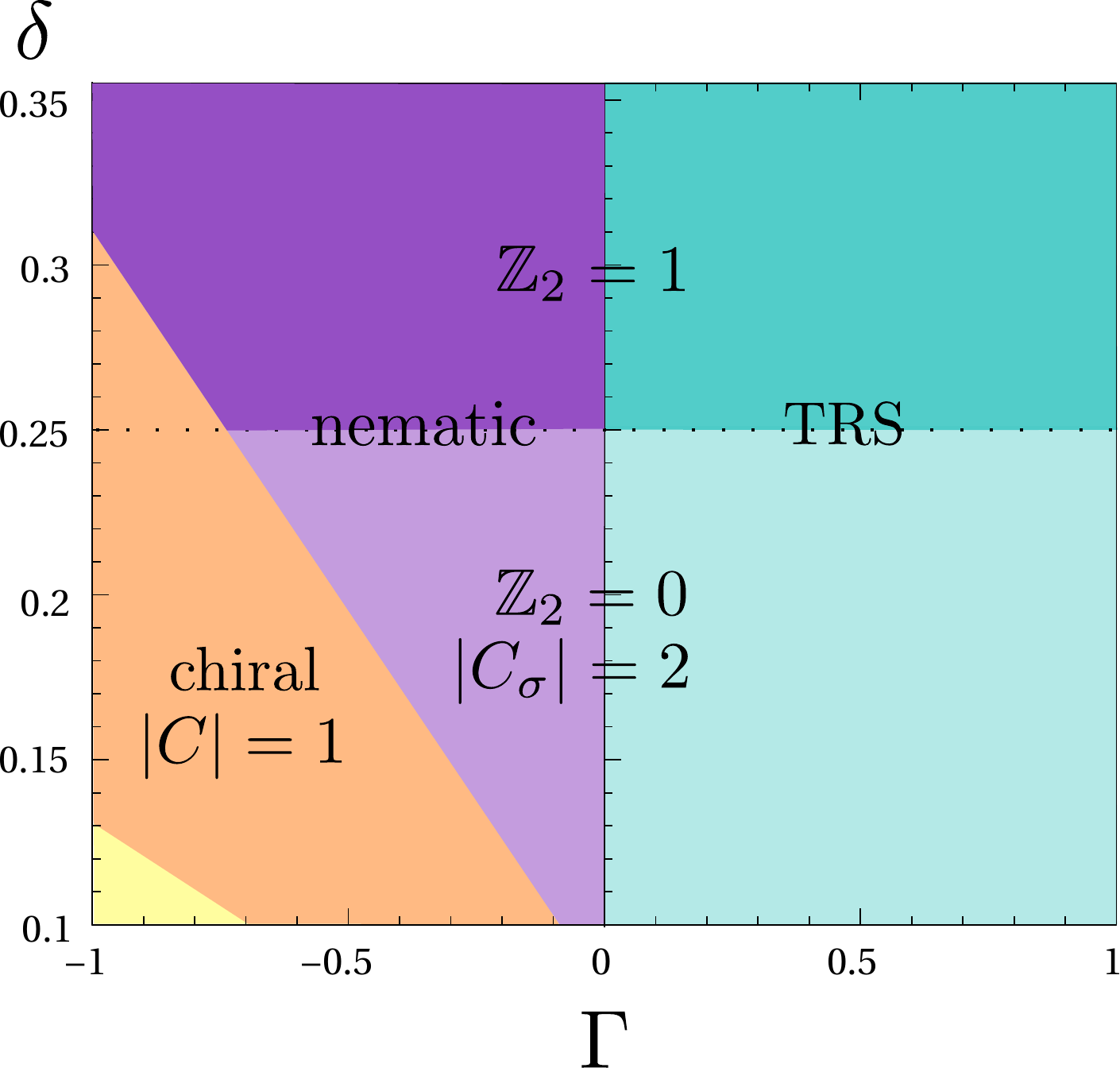}
\caption{Phase diagram of the triplet order parameters for $K=-t$ for positive and negative values of the symmetric off-diagonal exchange term $\Gamma$ as a function of doping $\delta$. For $\Gamma>0$, the time-reversal symmetric solution $\vc{d}_{\mathrm{TRS}}$ (cyan) is stable at all doping levels. Beneath $\delta=0.25$ (dashed line), it hosts a symmetry protected topological phase, while above the stronger $\mathbb{Z}_2$ invariant becomes non-trivial. Three different solutions appear for $\Gamma<0$. At large doping, the superconducting order $\vc{d}_{\mathrm{nematic}}$ (purple) breaks the $C_3$ symmetry, but is topologically equivalent to $\vc{d}_{\mathrm{TRS}}$, both above and below $\delta =0.25$. In the orange region, an order parameter $\vc{d}_{\mathrm{chiral}}$, breaking time-reversal symmetry and classified by a non-zero Chern number, appears. The solution in the yellow region at low doping mixes two irreps. \label{fig:negregion}}
\end{figure}

Starting by excluding the Heisenberg interaction $J$, we find triplet superconducting states throughout the full parameter space. This is illustrated in Fig.~\ref{fig:negregion}, where we plot the $\Gamma-\delta$ phase diagram. As seen, we find multiple different topological phases, breaking symmetries ranging from time-reversal symmetry, to crystalline symmetries by forming a nematic state. Very interestingly, just changing the sign of $\Gamma$ results in very different superconducting phases, which means that $\Gamma = 0$ is exactly at a phase transition for all doping levels.
In total, we identify four different regions, which we discuss in detail below.

\subsection{Time-reversal symmetry breaking states\label{sec:TRSB}}

We first identify a chiral, time-reversal symmetry breaking, odd-parity solution appearing at intermediate doping levels and $\Gamma<0$, see orange region in the phase diagram in Fig.~\ref{fig:negregion}. This is the only time-reversal symmetry breaking solution that we find.
The order parameter takes the form
\begin{align}
\label{eq:chiral}
\vc{d}_{\mathrm{chiral}} &= \eta
	\left(
		\begin{matrix}
			0 & 1 & \e{\pm \I 2 \pi /3} \\
			\e{\mp \I 2 \pi /3} & 0 & \e{\pm \I 2 \pi /3} \\
			\e{\mp \I 2 \pi /3} & 1 & 0
		\end{matrix}
	\right) \\
&= \eta
	\left\{
		\frac{1}{2}
		\left(
			\begin{matrix}
				0 & 2 & -1 \\
				-1 & 0 & -1 \\
				-1 & 2 & 0
			\end{matrix}
		\right)
		\pm \I \frac{\sqrt{3}}{2}
		\left(
			\begin{matrix}
				0 & 0 & 1 \\
				-1 & 0 & 1 \\
				-1 & 0 & 0
			\end{matrix}
		\right)
	\right\}\nonumber\\
&= \eta
	\left(
		\frac{1}{2}(\vc{d}_{\mathrm{E_{u},3}}+\vc{d}_{\mathrm{E_{u},4}}-2\vc{d}_{\mathrm{E_{u},5}}+\vc{d}_{\mathrm{E_{u},6}})
	\right. \nonumber\\
&
	\left.
	\pm \I \frac{\sqrt{3}}{2}(\vc{d}_{\mathrm{E_{u},4}}-\vc{d}_{\mathrm{E_{u},3}}+\vc{d}_{\mathrm{E_{u},6}})
	\right),\nonumber
\end{align}
where the to solutions $\pm$ are degenerate and of opposite chirality. We note directly that this order parameter is non-unitary as $\vc{d}_{\mathrm{chiral}}* \neq \vc{d}_{\mathrm{chiral}}$.\cite{Sigrist1991} Moreover, since the solution breaks time-reversal symmetry, but is still intrinsically particle-hole symmetric, it belongs to the Altland-Zirnbauer class D and is thus classified by a Chern number $C$ in 2D.\cite{Schnyder2008}

\begin{figure}[tb]
\includegraphics[scale=.4]{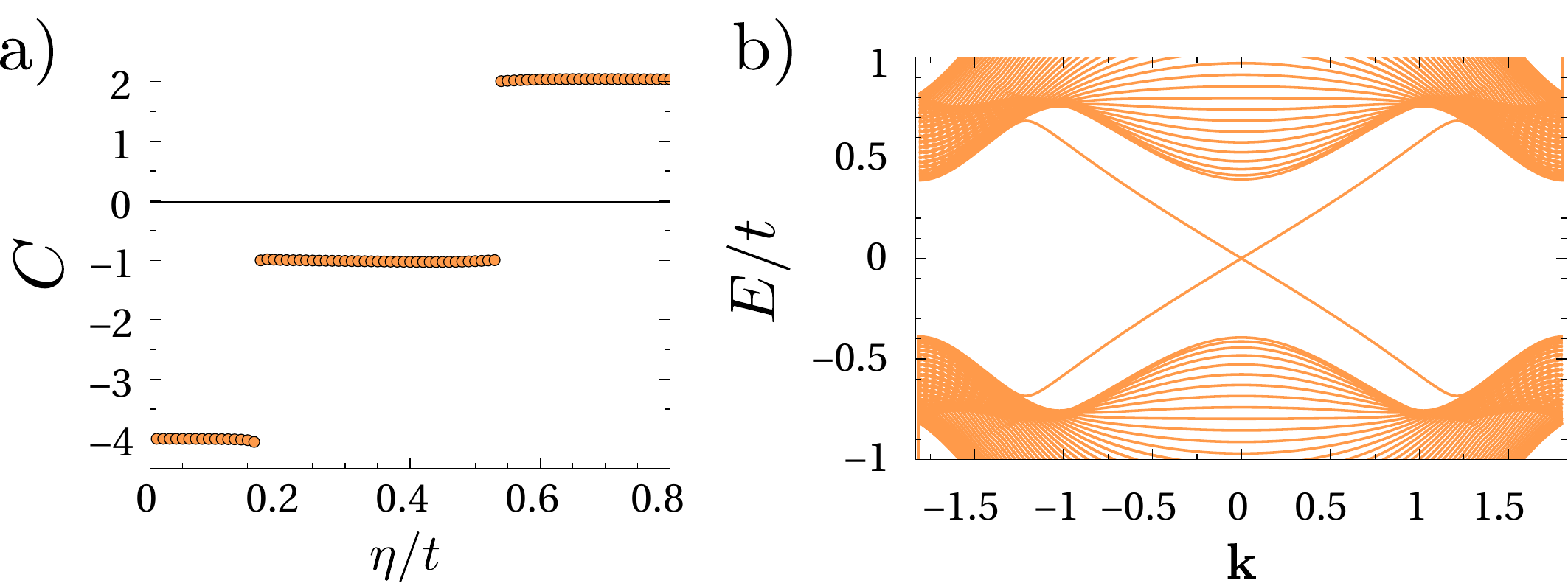}
\caption{Properties of the chiral superconducting state $\vc{d}_{\mathrm{chiral}}$ found at intermediate doping levels at $\Gamma<0$. (a) Non-self-consistent value of the Chern number $C$ for different order parameter strength $\eta$ at doping $\delta = 0.14$. Self-consistency only realizes the $C\pm1$ state. (b) Energy spectrum on a zigzag nanoribbon for $\eta=0.4t$ and $\delta=0.14$. There is one chiral edge state on each open boundary. \label{fig:Chernwithstrength}}
\end{figure}

We calculate the Chern number for various values of the order parameter strength $\eta$ in Eq.~\eqref{eq:chiral} using the numerical algorithm developed in Ref.~\onlinecite{Fukui2005} and present the results in Fig.~\ref{fig:Chernwithstrength}(a).
For all values of the order parameter strength $\eta$ and doping level $\delta$, the resulting Chern number is nonzero and the solution thus topologically non-trivial.
In particular, we find that the Chern number evolves with $\eta$. This is illustrated in Fig.~\ref{fig:Chernwithstrength}(a), which shows the Chern number $C$ calculated at $\delta \approx 0.14$ for one of the two degenerate solutions.
At weak pairing, with $\eta \ll t$, the Chern number is even and large, $C=-4$. Upon reaching a threshold $\eta_{c1}$, a topological phase transition occurs into a state with Chern number $C=-1$. At even higher values $\eta>\eta_{c2}$, another transition takes place into a phase with $C=2$. The other degenerate solution always yields a Chern number with opposite sign. At the critical values of $\eta$, at which the topological invariant changes, we find that the band gap closes at the $M$-point of the Brillouin zone. The exact values of $\eta$ for these  phase transitions depend on the doping level.
Due to the bulk-boundary correspondence, the finite Chern number in the chiral state is manifested in $|C|$ number of surface states on each open boundary. In Fig.~\ref{fig:Chernwithstrength}(b) we illustrate this for $\eta=0.4t$ and $\delta=0.14$, where $|C|=1$, by presenting the energy spectrum of a strip of honeycomb lattice with zigzag edges. There are two states crossing the bulk energy gap, which correspond to one chiral edge state on each zigzag edge.

While the chiral state in Eq.~\eqref{eq:chiral} is topologically non-trivial for any, even arbitrarily small, values of $\eta$, we find that only the solution with $C=\mp1$ is actually realized in our self-consistent calculations. The boundaries of the orange region in Fig.~\ref{fig:negregion} correspond exactly to the lines were the order parameter strength $\eta$, calculated from the self-consistency equations, reaches the critical values $\eta_{c1}$ and $\eta_{c2}$, respectively.
Because the self-consistently calculated strength of the order parameter depends on the exact values of the interaction parameters, the boundary of the orange region in Fig.~\ref{fig:negregion} depends rather strongly on the interaction strength. Repeating the calculations with smaller ratios of $K/t$ and $\Gamma/t$, we find that the chiral solution is still realized, but shifted to smaller doping values.

\begin{figure}[tb]
\includegraphics[width=.4\textwidth]{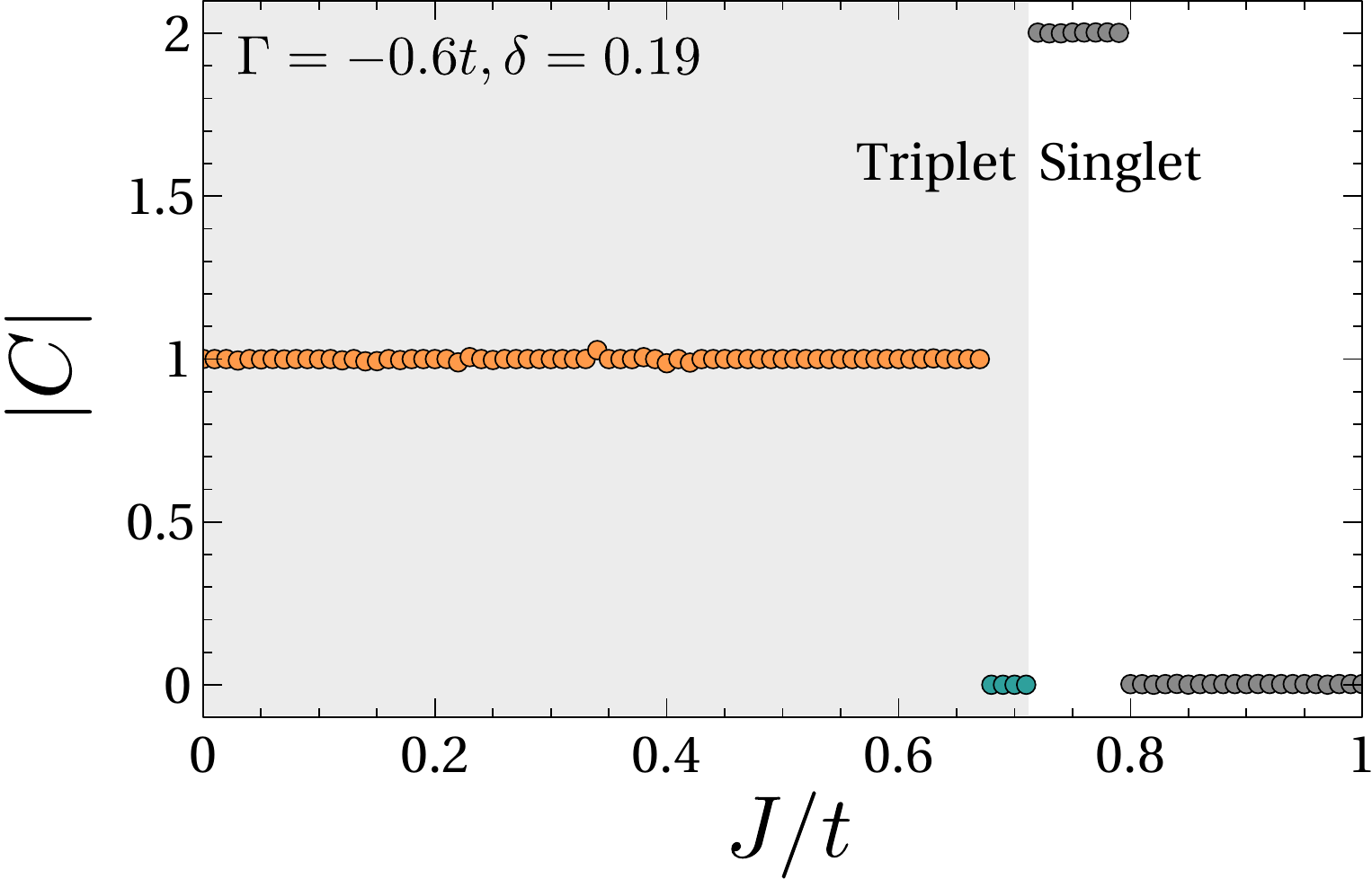}
\caption{Absolute value of the Chern number $C$ of the superconducting state obtained self-consistently at $\delta=0.19$, $K=-t$, and $\Gamma=-0.6t$, as a function of the Heisenberg interaction $0<J<t$. Grey shading marks the region of dominant triplet pairing. The chiral solution with $|C| = 1$ is dominant up to high values of $J$. Upon transitioning into the singlet state, a chiral $d$-wave solution with $|C| = 2$ is obtained until an extended $s$-wave state is stabilized at large values of $J$. A small intermediate region is also possibly present with a topologically trivial triplet state.
\label{fig:ChernwithJ}}
\end{figure}
When we also allow for a finite strength of the Heisenberg coupling $J$, the chiral spin-triplet state remains very stable. In Fig.~\ref{fig:ChernwithJ} we show the evolution of the absolute value of the Chern number calculated from the self-consistent solution obtained for $\delta = 0.19$, $\Gamma = -0.6t$, and $K=-t$ for increasing values of $J$. We find that the chiral solution with $|C| = 1$ is stable up to $J\approx 0.65t$. At higher $J$, there is a transition to a spin-singlet chiral $d$-wave state with $C=\pm2$ and eventually also into a singlet extended $s$-wave state, which is topologically trivial.
Shortly before the spin-singlet superconducting states are becoming favored, we find a small, and possibly unstable, intermediate region with a time-reversal symmetric spin-triplet solution with $|C|=0$. The symmetry of this transition region corresponds to the nematic phase discussed in Sec.~\ref{sec:TRSGneg}.
The transition to the chiral $d$-wave state is analogous to the case of $\Gamma=0$,\cite{Hyart2012} only the exact value of the transition is here slightly increased relative to the $\Gamma=0$ case. The transitions between different spin-singlet states have also been reported for the pure Heisenberg interaction.\cite{BlackSchaffer2007, Lothman2014} We can therefore conclude that the time-reversal symmetry breaking chiral triplet state is stable even in the presence of substantial Heisenberg coupling $J$.

\subsection{Time-reversal symmetric states}
The remaining phases in the $\Gamma-\delta$ phase diagram in Fig.~\ref{fig:negregion} all preserve time-reversal symmetry, while keeping an odd-parity due to their spin-triplet nature.
The different phases can be grouped together by the parameter ranges in which they occur. In particular, we separate the analysis according to the sign of $\Gamma$ in the phase diagram. However, before proceeding, we  first review some previous results in the case of $\Gamma = 0$, which are relevant for our subsequent discussion of the finite $\Gamma$ phases.

\subsubsection{Review of $\Gamma=0$\label{sec:G0}}
In the absence of the off-diagonal exchange $\Gamma$, the remaining interactions $K$ and $J$ do not mix the three different triplet components. The individual $\vc{d}$-vector components can then be classified without taking into account the spin-orbit coupling.\cite{Hyart2012} A symmetry analysis in this case yields three different irrep basis vectors for each component, see Appendix~\ref{sec:woSOC} for more details.
Employing the same mean-field approach as used here, Hyart {\it et al.}\cite{Hyart2012} found that each $\vc{d}$-vector component transforms according to the 2D $E_{u}$ irrep for all studied doping levels, as long as $J < J_c \approx \frac{1}{2} K$.
Moreover, the three $\vc{d}$-vector components were always found to be degenerate, forming in total four degenerate linear combinations with different directions of the $\vc{d}$-vector. In band space, these degenerate solutions all correspond to time-reversal symmetric, completely gapped odd-parity or $p$-wave states.

From the viewpoint of topological protection, all these $\Gamma=0$ solutions were found to exhibit a transition from a $\mathbb{Z}_2$ trivial to non-trivial state upon doping across the van Hove singularity at $\delta=0.25$, represented by the dashed line in Fig.~\ref{fig:negregion}.\cite{Hyart2012}
This transition is based on a weak coupling result: The $\mathbb{Z}_2$ invariant of a fully gapped, odd-parity superconducting order in the presence of inversion symmetry is given by the parity of the number of time-reversal invariant momentum (TRIM) points in the Brillouin zone enclosed by the normal-state Fermi surface.\cite{Sato2009,Sato2010}
In the case of the honeycomb lattice, the Fermi surface encloses the two $K$ and $K'$ points at doping below $\delta_c = 0.25$, such that $\mathbb{Z}_2 = 0$. At $\delta_c$, the Fermi surface undergoes a Lifshitz transition and after that only encloses the $\Gamma$ point, such that $\mathbb{Z}_2 = 1$.

However, even at doping levels below the van Hove singularity the system was found to be in a symmetry protected topological phase, since spin rotation symmetry of the kinetic part allows to write the full Hamiltonian in a block-diagonal form.\cite{Hyart2014} These individual blocks each have a non-vanishing spin Chern number $C_\sigma = \pm2$, which leads to the appearance of a pair of Dirac cones on an open boundary. As we will discuss in Section~\ref{sec:hopping}, this symmetry protection is however broken when the appropriate spin-orbit coupling terms is also included in the kinetic energy and not just in the Kitaev exchange.

\subsubsection{$\Gamma > 0$\label{sec:TRSGpos}}
Having reviewed the superconducting phase present at $\Gamma =0$, let us immediately turn to including a positive $\Gamma$. Here, we find the same superconducting order for all studied doping levels, cyan colored in Fig.~\ref{fig:negregion}. This order transforms according to $\vc{d}_{\mathrm{TRS}} = \eta \vc{d}_{\mathrm{A_{1u}}}$ given in Eq.~\eqref{eq:symclass}. It corresponds to one of the degenerate linear combinations present without the symmetric off-diagonal exchange term $\Gamma$, with the $\vc{d}$-vector now locked to point perpendicular to the honeycomb plane.

To understand why only this particular linear combination appears as the solution for $\Gamma >0$, we have to look into the definition of the spin coordinate system for the interaction Hamiltonian. The Kitaev exchange appears from virtual hopping processes between $j_{\mathrm{eff}} = \frac{1}{2}$ states on edge sharing oxygen octahedra, as illustrated in Fig.~\ref{fig:bonds}. Due to interference between different virtual hopping paths in the plane of the shared edge and the honeycomb lattice ions, only the spin component perpendicular to that plane interacts in the Kitaev exchange. For example, in the case of the $x$-bond in Fig.~\ref{fig:bonds}, this is set to be the $S^x$ axis. However, a different, downward pointing, spin coordiante axis $\tilde{S^x}= -S^x$ could have been defined. The Kitaev exchange $K S^x_i S^x_j$ is symmetric under such a change of basis.
Similarly, on the other two nearest-neighbor bonds, there is also always a choice of coordinate axis up or down for the spin component involved in the Kitaev exchange, without changing the interaction. Three bonds, each with a choice of up and down, give eight different spin coordinate systems that produce the same Kitaev exchange, but only four of them are right-handed, as needed. The resulting four choices are
\begin{align}
\label{eq:quantaxes}
&(S^x, S^y, S^z), & &
(S^x, -S^y, -S^z),\\
&(-S^x, S^y, -S^z), & &
(-S^x, -S^y, S^z),\nonumber
\end{align}
when written in terms of the spin axes used in Fig.~\ref{fig:bonds}.

This symmetry of the interaction is broken, as soon as a finite off-diagonal exchange $\Gamma$ is included. The extended Kitaev-Heisenberg Hamiltonian is no longer unchanged between the different choices of spin coordinate systems in Eq.~\eqref{eq:quantaxes}, because the off-diagonal exchange includes terms of the form $S^x_i S^y_j$. In our choice of coordinate system, all off-diagonal terms come with a positive sign in $H_{\mathrm{KJ\Gamma}}$, see Eq.~\eqref{eq:extKHHam}:
\begin{align*}
\Gamma \left(S^y_i S^z_1 + S^z_i S^y_1 + S^x_i S^z_2 + S^z_i S^x_2 + S^x_i S^y_3 + S^y_i S^x_3 \right)
\end{align*}
Using e.g.~the second coordinate system in Eq.~\eqref{eq:quantaxes} would transform $\Gamma S^x_i S^y_j \rightarrow -\Gamma S^x_i S^y_j$ and thus does not leave the interaction invariant. Hence, the parametrization of the off-diagonal exchange fixes the spin coordinate system and breaks the four-fold symmetry of the interaction at $\Gamma=0$.
The symmetry classification in Section~\ref{sec:symanal} was performed using the first set of quantization axis in Eq.~\eqref{eq:quantaxes}. If writing down the Hamiltonian using any of the other three sets, the symmetry classification needs to be adjusted. For example, the axis perpendicular to the honeycomb plane, that was necessary for the the $C_3$, or $\frac{2\pi}{3}$ rotation, will no longer correspond to the $(1,1,1)$ axis in the new choice of coordinates. Performing the symmetry analysis in the four different coordinate systems, the basis function of the $A_{1u}$ irrep will always correspond to one of the other four linear combinations found to be degenerate at $\Gamma=0$.

The overall conclusion is that including a positive off-diagonal exchange $\Gamma$ locks the $\vc{d}$-vector perpendicular to the honeycomb plane. This process strongly singles out one of the four degenerate states at $\Gamma =0$ as the ground state.
Since the off-diagonal exchange $\Gamma>0$ does not change the symmetry of the superconducting order, the topological classification by Hyart {\it et al.}\cite{Hyart2012,Hyart2014} carries over.
However, we find that including a finite $\Gamma$ enhances the magnitude of the superconducting order. Comparing for example the amplitude $\eta$ at $\Gamma=0$ with that at $\Gamma=t$ for a fixed doping, there is an increase by up to an order of magnitude.
Finally, including a finite Heisenberg interaction leads to a transition to a chiral $d$-wave and eventually an extended $s$-wave singlet superconducting order. We find that a finite $\Gamma>0$ slightly increases the critical value $J_c$ at which the transition occurs.

\subsubsection{$\Gamma < 0$\label{sec:TRSGneg}}
For negative values of $\Gamma$, we also observe time-reversal symmetric solutions in addition to the time-reversal breaking superconducting order already discussed in Sec.~\ref{sec:TRSB}. The dominating phase is obtained at intermediate to high doping levels, see purple region in Fig.~\ref{fig:negregion}. We identify several parameterizations which are degenerate in energy in this region. They can all be expressed as purely real linear combinations of the basis vectors of the $E_{u}$ irrep
\begin{align}
\vc{d}_{\mathrm{nematic}} = \sum_i a_i \vc{d}_{\mathrm{E_{u}},i}.
\end{align}
The coefficients of the first two basis functions of the $E_u$ irrep, $a_1$ and $a_2$, are always at least an order of magnitude smaller than the other coefficients. This results in the three $\vc{d}$-vector components being of different magnitude. One particular example for these solutions is
\begin{align}
\label{eq:TRSprime}
\vc{d}_{\mathrm{nematic'}} = \left(
\begin{matrix}
0 & 0 & 1 \\
-1 & 0 & 1 \\
-1 & 0 & 0
\end{matrix}\right) = (\vc{d}_{\mathrm{E_{u},6}}+\vc{d}_{\mathrm{E_{u},4}}-\vc{d}_{\mathrm{E_{u},3}}).
\end{align}

\begin{figure}[tb]
\includegraphics[scale=.27]{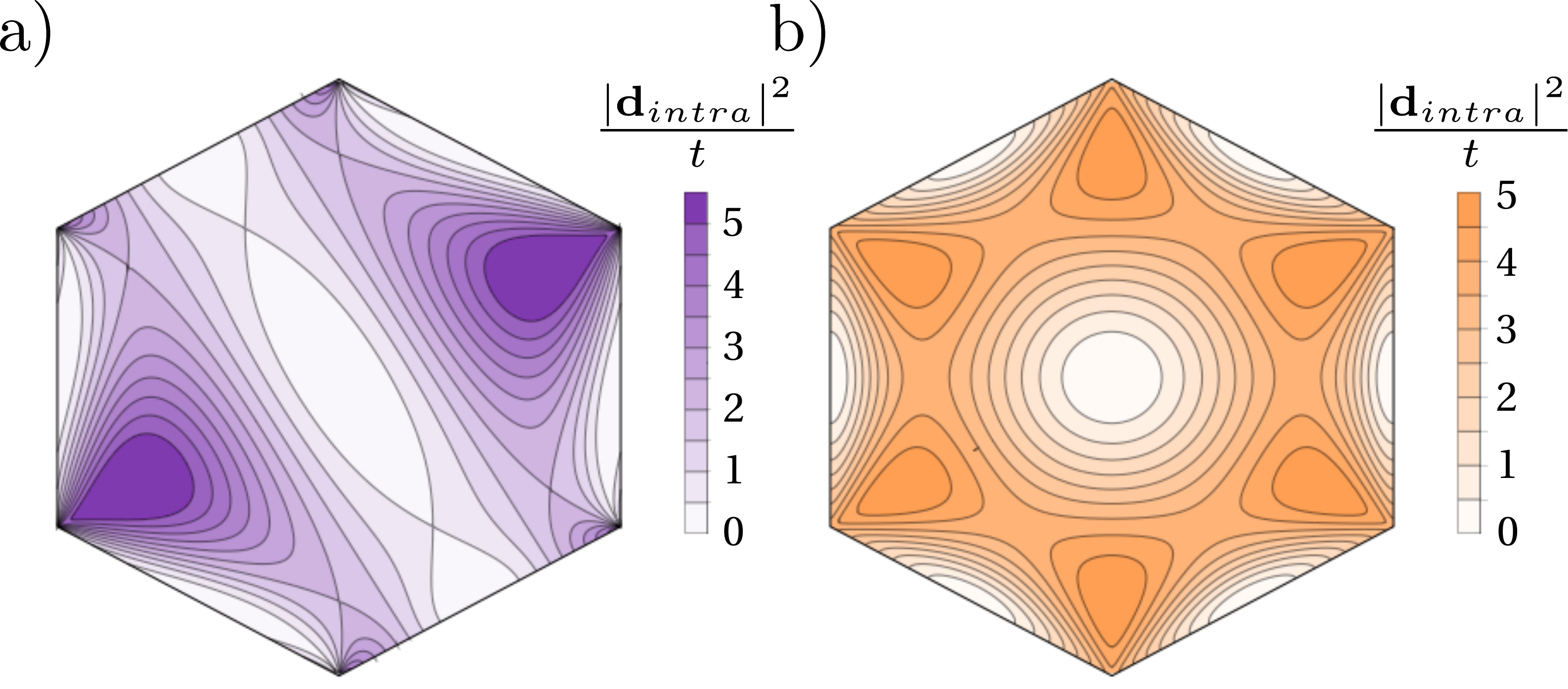}
\caption{Square of the absolute value of the intraband triplet order parameter $|\vc{d}_{\mathrm{intra}}(\vc{k})|^2$ in the first Brillouin zone for (a) the time-reversal symmetric nematic order and (b) the chiral time-reversal breaking order, both present at $\Gamma<0$. Breaking of the $C_3$ symmetry is clearly visible in (a).
\label{fig:intraband}}
\end{figure}

To gain more insights into this state, we transform the order parameter into the band picture, where the normal state Hamiltonian $H_\mathrm{k}$ is diagonal. Then the superconducting pairing can be divided into intra- and interband contributions. The corresponding transformations are explained in detail in Appendix~\ref{sec:appband}. Because we expect the intraband contribution to be dominant at the doping levels relevant for this state, we plot the square of the absolute value of the intraband pairing $|\vc{d}_{\mathrm{intra}}(\vc{k})|^2$ in the first Brillouin zone in Fig.~\ref{fig:intraband}. Panel (a) clearly shows that the $C_3$ symmetry is broken for this order, manifesting a clear nematicity in this phase. In contrast, Fig.~\ref{fig:intraband}(b) displays the intraband order parameter $\vc{d}_{\mathrm{chiral}}$ for the chiral time-reversal symmetry breaking states, which retains the full rotational symmetry of the honeycomb lattice.

We have also confirmed that the subdominant interband order parameter has the same spatial symmetries as the intraband pairing.
Finally, numerical calculations of the energy band gap also show a clear nematicity by breaking the $C_3$ symmetry. Thus the dominating time-reversal symmetric solution for $\Gamma<0$ is a nematic state, with the nematicity clearly distinguishable in experiments measuring the energy gap. The remaining spatial symmetry of the order parameter in the nematic states corresponds to a $C_2'$ rotation. This directly explains the energy degeneracy of several solutions in this phase,\footnote{Numerically we find very near degenerate states, as a perfect energy degeneracy is hard to achieve due to finite sampling in the Brillouin zone.} as they correspond to different spontaneous choices of symmetry axis for the $C_2'$ symmetry. Finding several degenerate solutions is thus a natural consequence of the nematic order.
In fact, the nematicity of this time-reversal symmetric order is in agreement with very recent results showing that odd-parity superconducting orders classified by a 2D irrep in 3D superconductors with strong spin-orbit coupling must either have chiral or nematic symmetry.\cite{Venderbos2016} Assuming these results are extendable to 2D systems, the retained time-reversal symmetry of the purple region in Fig.~\ref{fig:negregion} implicates a nematic state, as we also find.

Even though the nematic phase has a different spatial symmetry than the $\vc{d}_{\mathrm{TRS}}$ state observed at $\Gamma\geq0$, the topological classification is actually completely equivalent. Both phases preserve time-reversal symmetry, have an order parameter with odd-parity, and are gapless. Thus the  value of the $\mathbb{Z}_2$ is fully determined by the normal state Fermi surface in the weak-coupling limit. This leads to a nontrivial $\mathbb{Z}_2 =1$ for doping levels above the Lifshitz transition at $\delta_c =0.25$, but $\mathbb{Z}_2 =0$ below. But, as also discussed in Section~\ref{sec:TRSGpos}, at $\delta<0.25$ there is an additional symmetry protection through $C_\sigma = \pm 2$.
Furthermore, we find that the influence of the Heisenberg interaction $J$ on the nematic order is also very similar to the $\vc{d}_{\mathrm{TRS}}$ case, with a transition to the singlet chiral $d$-wave state as discussed in Section~\ref{sec:TRSGpos}, but only for large $J$.

Finally, at low doping, we observe a phase, where the order parameter mixes the $E_u$ and $A_{2u}$ irreps, but remains invariant under time-reversal symmetry. This region is small and marked in yellow in the phase diagram in Fig.~\ref{fig:negregion}.  The topological classification is equivalent to the time-reversal symmetric solutions discussed earlier. However, because this solution is found so close to the limits of our theory, it is likely an artifact of the artificially strong coupling. In fact, when choosing smaller interaction parameters, this region is quickly pushed below the $\delta=0.1$ boundary. Therefore we do not analyze this solution further.

\section{Spin-orbit Coupling\label{sec:hopping}}
Above we have classified, both topologically and symmetry-wise, all different superconducting states appearing in the extended Kitaev-Heisenberg model using a natural incorporation of doping effects given by the total Hamiltonian $H = H_\mathrm{k} + H_\mathrm{\Delta}$.
In this section, we discuss the influence of an additional spin-orbit coupled hopping term in the kinetic energy.
This Kane-Mele like spin-orbit coupling was first identified in Na$_2$IrO$_3$,\cite{Shitade2009} where it stems from a second nearest-neighbor hopping pathway between unlike $t_{2g}$ orbitals via the Na ions in the center of the honeycombs.\cite{Mazin2012,Foyevtsova2013} This hopping process has later also been found to be present in the other materials realizing the extended Kitaev-Heisenberg model.\cite{Winter2016} It takes the form
\begin{align}
\label{eq:HSO}
H_{\mathrm{SO}} = \I t' \! \! \sum_{\braket{\braket{i,j}},\sigma,\bar{\sigma},o} \! \! \left( c^\dagger_{i,\sigma,o} \sigma^{\gamma(i,j)}_{\sigma,\bar{\sigma}} c_{j,\bar{\sigma},o} + {\rm H.c.} \right),
\end{align}
where the Pauli matrix $\sigma^{\gamma(i,j)}$ involved in the hopping process depends on the link to the  second nearest-neighbor $\braket{\braket{i,j}}$, similar to the way the Kitaev exchange involves different spin components.

\begin{figure}[tb]
\includegraphics[scale=.33]{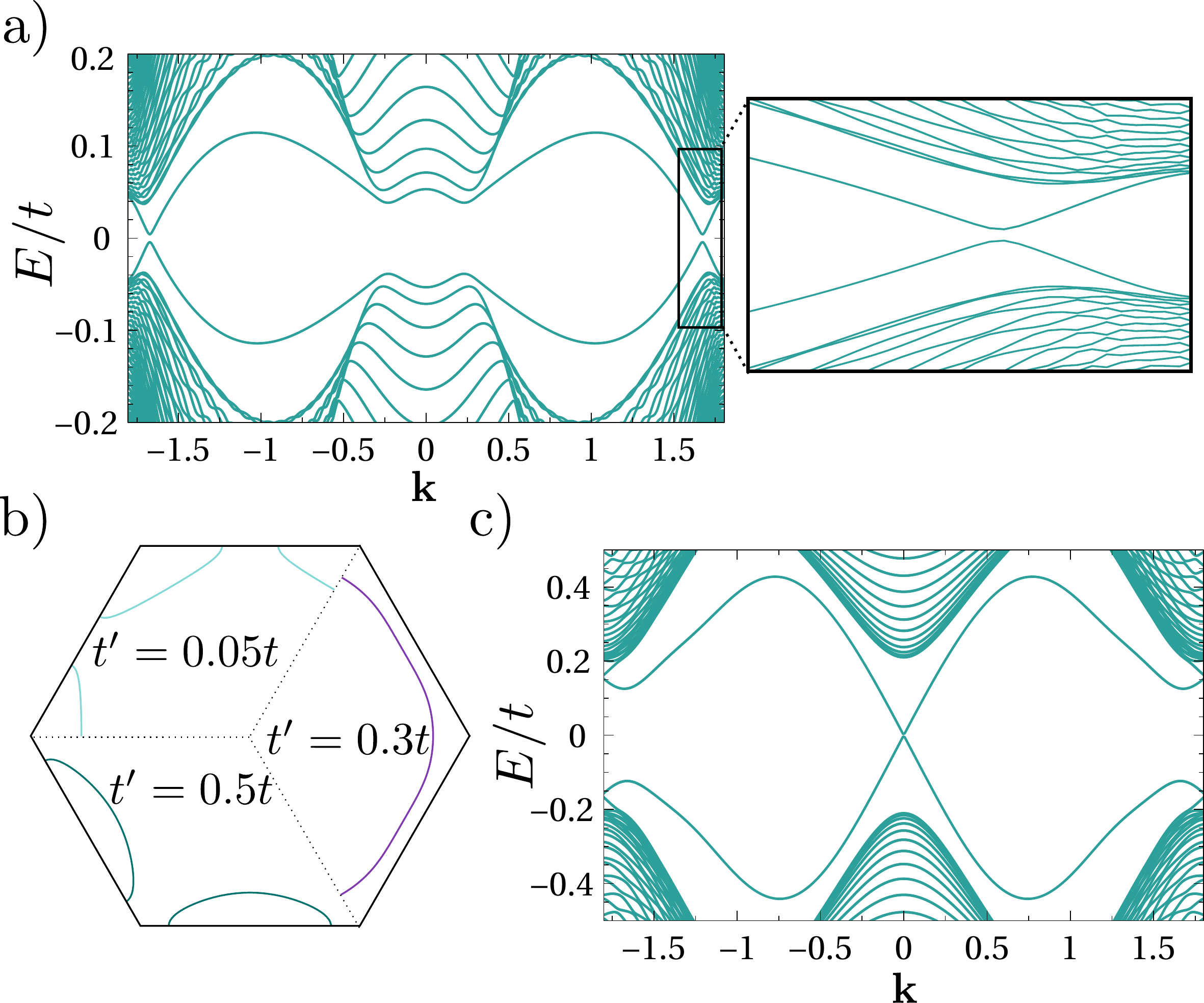}
\caption{Influence of the spin-orbit hopping term $H_{\mathrm{SO}}$ on the topological classification of the time-reversal symmetric superconducting states. (a) Edge state spectrum of $\vc{d}_{\mathrm{TRS}}$ including $t'=0.02$ at $\delta=0.2$. The subgap states are gapped, see zoom-in inset. (b) Evolution of the normal-state Fermi surface at constant filling $\delta=0.2$ for different values of the spin-orbit hopping $t'$. (c) Edge state spectrum of $\vc{d}_{\mathrm{TRS}}$ at $\Gamma=0.5$, $\delta=0.2$, and $t'=0.3$. The system is in a $\mathbb{Z}_2$ non-trivial state driven by $t'$ despite $\delta<0.25$.
 \label{fig:FSwithShitade}}
\end{figure}

Adding $H_{\mathrm{SO}}$ to the kinetic energy does not allow the decomposition of the Hamiltonian into diagonal blocks anymore, since this term breaks the $SU(2)$ symmetry of the kinetic Hamiltonian. Thus, the symmetry protected phase with $C_\sigma = \pm 2$ discussed previously in Section~\ref{sec:G0}, and found in all the time-reversal symmetric states at $\delta<0.25$, cannot be viewed as a realistic prediction for materials described by the extended Kitaev-Heisenberg model.
In Fig.~\ref{fig:FSwithShitade}(a), we support this statement by presenting the edge state spectrum of a zigzag strip with open boundary conditions. The superconducting order corresponds to $\vc{d}_{\mathrm{TRS}}$ and the doping level is $\delta=0.2$, i.e.~well within the presumed symmetry protected topological phase. Adding the spin-orbit hopping term $H_{\mathrm{SO}}$ to the kinetic energy immediately lifts the symmetry protection of the edge states, leading to a gapping of both Dirac cones, as clearly seen in the inset. This shows explicitly that the spin-orbit induced kinetic energy breaks the symmetry needed to enter the $C_\sigma = \pm 2$ state.

To study more systematically the influence of the spin-orbit hopping term in Eq.~\eqref{eq:HSO} on the time-reversal symmetric superconducting states, we self-consistently calculate the order parameter for $\Gamma=0.5$ at the doping level $\delta = 0.2$ and tuning $t'$ up to $0.5t$. The symmetry of the superconducting solution does not change compared to $t'=0$ for this range of values, realizing $\vc{d}_{\mathrm{TRS}}$ throughout. However, the effect on the band structure leads to a new topological transition. The change in the dispersion caused by $t'$ is accompanied by two Lifshitz transitions, illustrated in Fig.~\ref{fig:FSwithShitade}(b). The Fermi pockets around the $K$ and $K'$ points, that make up the Fermi surface at doping $\delta<0.25$ in the absence of $t'$, first give way for a pocket around the $\Gamma$ point, before the Fermi surface eventually forms pockets around the $M$ points for increasing values of $t'$. During the first transition the number of TRIM points in the Brillouin zone points enclosed by the Fermi surface changes from an even to an odd number. This leads to a topological phase transition into a $\mathbb{Z}_2$ non-trivial state.\cite{Sato2009,Sato2010} This is the same topological argument as used across the doping-driven Lifshitz transition at $\delta_c =0.25$, but in this case it takes place at fixed doping and is instead driven by $t'$. We illustrate this topological states in Figure~\ref{fig:FSwithShitade}(c), which shows the spectrum of a zigzag strip with open boundary conditions for the $\vc{d}_{\mathrm{TRS}}$ solution at $\Gamma=0.5$, $\delta=0.2$ and $t'=0.3$. The topologically protected subgap states appearing on the edges of the ribbon due to the change into the non-trivial $\mathbb{Z}_2$ state after the first $t'$-driven Lifshitz transition are clearly visible.
While we presented these results only for the time-reversal symmetric order stable at $\Gamma\geq0$, the same arguments also directly apply to the nematic order, since it is topologically equivalent to $\vc{d}_{\mathrm{TRS}}$.

The chiral state $\vc{d}_{\mathrm{chiral}}$ is not affected by the inclusion of the additional spin-orbit hopping. We have performed self-consistent calculations again at $\delta=0.2$, i.e.~deep within the region where the chiral order is stabilized, and find that the spin-orbit kinetic energy does not affect the symmetry of the order parameter.
The change of the Fermi surface with increasing $t'$ does also not lead to a change in the topological invariant.
When time-reversal symmetry is broken, the parity of the Chern number in the weak-coupling limit is also determined by the number of TRIM points enclosed by the Fermi surface, however spin-degenerate bands now have to counted individually.\cite{Sato2010} This leads to an even Chern number at weak coupling for any Fermi surface of the spin-degenerate normal state on the honeycomb lattice. As a consequence, the $t'$-driven Lifshitz transition does not influence the topology of the chiral superconducting state.
Nonzero values of $t'$ will, however, somewhat change $\eta_{c1}$ and $\eta_{c2}$ governing the transition in an out of the $C =1$ phase, and thus the transition from the $\vc{d}_{\mathrm{chiral}}$ to the $\vc{d}_{\mathrm{nematic}}$ phase.  We thus find that the boundaries of the time-reversal symmetry breaking region generally move to lower doping regions upon increasing $t'$.

To summarize, including a finite spin-orbit hopping term to appropriately account for spin-orbit effects also in the kinetic energy results in notable modifications of the topological classification of all time-reversal symmetric superconducting states in the extended Kitaev-Heisenberg model. The $C_\sigma = \pm 2$ symmetry found at $\delta<0.25$ for $t' =0$ in all time-reversal symmetric states is now formally broken, with gapped boundary states as a direct consequence.
However, a large magnitude $t'$ of the spin-orbit hopping term can drive a Lifshitz transition of the normal-state Fermi surface also below $\delta=0.25$, such that a $\mathbb{Z}_2$ non-trivial state topological state appears. This latter state is equivalent to the state found at $\delta>0.25$, but it is here driven by $t'$ and not the doping level.
The time-reversal broken phase is, however, not influenced by $t'$.

%
% -------------------------------------------------- %
% Conclusions
% -------------------------------------------------- %
\section{Concluding Remarks\label{sec:conclusion}}

We have studied the possible superconducting orders arising upon doping the extended Kitaev-Heisenberg model on the honeycomb lattice, while keeping the full periodicity of the lattice. We find that the previously ignored symmetric off-diagonal exchange term $\Gamma$ only influences the triplet superconducting channels and there causes a mixing  of the components of the $\vc{d}$-vector. This mixing is also necessarily manifested in the symmetry analysis that we perform, which becomes significantly more involved compared to if $\Gamma$ is ignored. As a consequence, we find multiple topological states, which are also very sensitive to the sign of $\Gamma$.

Performing self-consistent calculations, we map out the $\Gamma$-$\delta$ phase diagram, which displays several remarkable triplet superconducting phases.
A spin-triplet chiral solution of non-trivial topology and breaking time-reversal symmetry is realized for $\Gamma <0$ at intermediate doping levels. We calculate the Chern number and show that this phase hosts a single topological edge state on each open boundary.
Other chiral phases with different Chern numbers, but with the same order parameter symmetry, are in principle also possible, but are never stabilized in the self-consistent calculations.
Instead, we discover a competing spin-triplet state with nematic symmetry present for $\Gamma<0$ at higher doping levels. It retains time-reversal symmetry, but manifests itself by breaking the $C_3$ rotational symmetry down to $C_2'$ for both the order parameter and energy gap. This nematicity should be visible in angular resolved experiments, such as the upper critical field or NMR studies of the Knight shift.

We find that the boundary between the chiral and nematic phases depends strongly on the ratio between interaction strength and bandwidth, but always lies such that only the chiral state with $|C|=1$ is stabilized and is thus also heavily doping dependent. This opens for the possibility to tune between the chiral and nematic phases by simply changing the doping of the material. 
A competition between a chiral and nematic state arising for a 2D irrep in spin-orbit coupled, odd-parity superconductors has recently been discussed in general terms in 3D superconductors.\cite{Venderbos2016} Here we find a very similar situation but in a class of 2D materials. The reduced dimensionality results in both the chiral and the nematic phases being fully gapped, in contrast to nodal features proposed at the north/south poles in the 3D chiral state.
Expanding this work to analyze the boundary between the chiral and nematic states observed in the extended Kitaev-Heisenberg model could provide further insights into the remarkable connection we find between Chern number and stability for the chiral state.

While the $\Gamma<0$ phase diagram shows this intricate competition between chiral and nematic states, we find that $\Gamma>0$ stabilizes the spin-triplet superconducting symmetry previously obtained at $\Gamma=0$. However, including a finite $\Gamma>0$ results in the spin-orbit coupling locking the $\vc{d}$-vector perpendicular to the honeycomb plane, which lifts a fourfold degeneracy present when ignoring $\Gamma$. The superconducting state at $\Gamma>0$ retains the full rotational symmetry of the lattice and is also time-reversal invariant. This means that there are clear experimental signatures differentiating between the superconducting orders stabilized by positive and negative values of $\Gamma$. In fact, the large sensitivity of the superconducting state to the sign of $\Gamma$ can be used as an accurate probe for determining the value of $\Gamma$ in these materials.
We also certify that all identified spin-triplet solutions remain stable under the inclusion of small to moderate Heisenberg interactions, as present in all known materials realizing the extended Kitaev-Heisenberg model, despite this term favoring a spin-singlet chiral $d$-wave state.

In addition, we study the influence of including a finite spin-orbit hopping term in the kinetic energy, that should be present in relevant materials. This term lifts the symmetry protection of the topological state of all time-reversal symmetric states below the critical doping of $\delta=0.25$. However, the same spin-orbit term can also drive a topological phase transition into a state with a non-zero $\mathbb{Z}_2$ invariant for the time-reversal symmetric phases, due to a Lifshitz transition at constant filling.
While the proposed spin-liquid ground state of the undoped extended Kitaev-Heisenberg model has recently generated great interest, our findings of a multitude of different topological and nematic superconducting states generated upon doping unveil an equally interesting superconducting phase diagram in these materials.

% -------------------------------------------------- %
% ACKNOWLEDGMENTS
% -------------------------------------------------- %
\begin{acknowledgments}
We are grateful to A.~Bouhon, T.~Hyart, and T. L\"othman for discussions. This work was supported by the Swedish Research Council (Vetenskapsr\aa det), the Swedish Foundation for Strategic Research (SSF), the G\"{o}ran Gustafsson Foundation, and the Knut and Alice Wallenberg Foundation through the Wallenberg Academy Fellows program.
\end{acknowledgments}

\appendix
\section{Symmetry classification without SOC \label{sec:woSOC}}

When considering spin and orbital degrees of freedom independently, the action of the symmetry operations in the point group leave the spin, and therefore the $\vc{d}$-vector, untouched. Then the three $\vc{d}$-vector components can be treated individually. This is the classification performed in Ref.~\onlinecite{Hyart2012}
In this case are three basis functions for each $\vc{d}$-vector component, one belonging to the $A_{1u}$ irrep and two to the $E_u$ irrep. The classification for each component is completely analogous, such that the resulting basis functions for each component in not normalized and not orthogonalized form are

\begin{align}
\label{eq:nonSOC}
\vc{d}^\gamma_{\mathrm{A_{1u}}} = \left(1, 1, 1\right),
& &
\vc{d}^\gamma_{\mathrm{E_{u},1}} = \left(-1 , 0 , 1\right),
& &
\vc{d}^\gamma_{\mathrm{E_{u},2}} = \left(-1 , 1 , 0\right),
\end{align}

for $\gamma=x,y,z$ and each of the three nearest-neighbor bonds. When grouping the basis function by irrep, there are thus three basis functions for the 1D $A_{1u}$ irrep and six basis functions in the 2D $E_u$ irrep.

\section{Band picture \label{sec:appband}}

The kinetic Hamiltonian $H_\mathrm{k}$ can be Fourier transformed to yield
\begin{align}
H_{\mathrm{k}} &= \! -\!\sum_{\vc{k},j,\sigma} t \e{\I \vc{k}\cdot\delta_j} \! \left( \! c^\dagger_{\vc{k},\sigma,a} c_{\vc{k},\sigma,b} + {\rm H.c.} \! \right) + \tilde{\mu} \sum_{\vc{k},\sigma,o} c^\dagger_{\vc{k},\sigma,o} c_{\vc{k},\sigma,o},
\end{align}
which is diagonalized by the unitary transformation
\begin{align}
\left(
\begin{matrix}
c_{\vc{k},\sigma,a} \\ c_{\vc{k},\sigma,b}
\end{matrix}\right)  &=
\frac{1}{\sqrt{2}}
\left(
\begin{matrix}
1 & 1\\
\e{-\I \phi_\vc{k}} & -\e{-\I \phi_\vc{k}}
\end{matrix}
\right)
\left(
\begin{matrix}
b_{\vc{k},\sigma,1}\\b_{\vc{k},\sigma,2}
\end{matrix}
\right).
\end{align}
Here, $b_{\vc{k},\sigma,l}$ annihilates an electron in band $l$, while $\phi_\vc{k} = \arg(\sum_j \e{\I \vc{k}\cdot\delta_j})$. Introducing the shorthand notation $\epsilon_\vc{k} = -t |\sum_j \e{\I \vc{k}\cdot\delta_j}|$, the kinetic part of the Hamiltonian in band space takes the diagonal form
\begin{align}
H_\mathrm{k} & = \sum_{\vc{k},\sigma} \left( (\epsilon_\vc{k} + \mu) b^\dagger_{\vc{k},\sigma,1} b_{\vc{k},\sigma,1} + (-\epsilon_\vc{k} + \mu) b^\dagger_{\vc{k},\sigma,2} b_{\vc{k},\sigma,2}\right).
\end{align}

The mean-field pairing Hamiltonian $H_\mathrm{\Delta}$ can likewise be transformed into this band basis. This has already been done for the singlet order parameter\cite{BlackSchaffer2007}, so here we focus on the triplet pairing terms, which transform as
\begin{widetext}
\begin{align}
\sum_{j} d^\alpha_{j} t^{\alpha\dagger}_{\vc{k},j}&= \sum_{\alpha} d^\alpha_{j} \frac{1}{\sqrt{2}} \sum_{\sigma,\bar{\sigma}} c^\dagger_{\vc{k},\sigma,a} c^\dagger_{-\vc{k},\bar{\sigma},b} \e{-\I \vc{k}\cdot\delta_j} \I \left( \mathbf{\sigma_y} \mathbf{\sigma_\alpha} \right)_{\sigma,\bar{\sigma}}\\
&= \sum_{j} d^\alpha_{j} \frac{1}{2\sqrt{2}} \sum_{b_1,b_2} b^\dagger_{\vc{k},\sigma,b_1} b^\dagger_{-\vc{k},\bar{\sigma},b_2} \e{-\I \vc{k}\cdot\delta_j} \e{\I \phi_\vc{k}} (\tau^z - \I \tau^y)_{b_1,b_2} \I \left( \mathbf{\sigma_y} \mathbf{\sigma_\alpha} \right)_{\sigma,\bar{\sigma}}\\
&= \sum_{j} d^\alpha_{j} \frac{1}{2\sqrt{2}} \sum_{b_1,b_2} \frac{1}{2}(b^\dagger_{\vc{k},\sigma,b_1} b^\dagger_{-\vc{k},\bar{\sigma},b_2}-b^\dagger_{-\vc{k},\bar{\sigma},b_2}b^\dagger_{\vc{k},\sigma,b_1}) \e{-\I \vc{k}\cdot\delta_j} \e{\I \phi_\vc{k}} (\tau^z - \I \tau^y)_{b_1,b_2} \I \left( \mathbf{\sigma_y} \mathbf{\sigma_\alpha} \right)_{\sigma,\bar{\sigma}}\\
&= \sum_{j} d^\alpha_{j} \frac{1}{2\sqrt{2}} \sum_{b_1,b_2} b^\dagger_{\vc{k},\sigma,b_1} b^\dagger_{-\vc{k},\bar{\sigma},b_2} \frac{1}{2}(\e{-\I \vc{k}\cdot\delta_j} \e{\I \phi_\vc{k}} + (-1)^{\delta_{b_1,b_2}} \e{\I \vc{k}\cdot\delta_j} \e{-\I \phi_\vc{k}}) (\tau^z - \I \tau^y)_{b_1,b_2} \I \left( \mathbf{\sigma_y} \mathbf{\sigma_\alpha} \right)_{\sigma,\bar{\sigma}}\\
&= \sum_{j} d^\alpha_{j} \frac{1}{2\sqrt{2}} \sum_{b_1,b_2} b^\dagger_{\vc{k},\sigma,b_1} b^\dagger_{-\vc{k},\bar{\sigma},b_2} (-\sin(\vc{k}\cdot\delta_j-\phi_\vc{k})\I \tau^z - \cos(\vc{k}\cdot\delta_j-\phi_\vc{k})\I \tau^y)_{b_1,b_2} \I \left( \mathbf{\sigma_y} \mathbf{\sigma_\alpha} \right)_{\sigma,\bar{\sigma}}.
\end{align}
\end{widetext}
The matrices $\tau^\gamma$ act on band space and allow us to introduce the intra- and interband triplet order parameters
\begin{align}
d^\alpha_{\mathrm{intra}} (\vc{k}) &= \sum_j d^\alpha_j \sin(\vc{k}\cdot\delta_j - \phi_\vc{k})\\
d^\alpha_{\mathrm{inter}} (\vc{k}) &= \sum_j d^\alpha_j \cos(\vc{k}\cdot\delta_j - \phi_\vc{k}).
\end{align}
Note that $\sin$ and $\cos$ terms are here exchanged in comparison to the singlet intra- and interband pairing. Diagonalizing the full Hamiltonian yields the quasiparticle spectrum, which, can be written as
\begin{align}
E(\vc{k}) &= \sqrt{\xi_1(\vc{k})^2 + \xi_2(\vc{k})^2 + |\vc{d}_{\mathrm{intra}}(\vc{k})|^2 + |\vc{d}_{\mathrm{inter}}(\vc{k})|^2 + M},
\end{align}
where $\xi_{1/2}(\vc{k}) = \pm \epsilon_{\vc{k}} + \mu$ is the dispersion of the two bands and the last term $M$ contains terms mixing intra- and interband pairing, and differing depending on whether the pairing is unitary or not.

% -------------------------------------------------- %
% BIBLIOGRAPHY:
% -------------------------------------------------- %
\bibliographystyle{apsrevmy}

\begin{thebibliography}{57}
\expandafter\ifx\csname natexlab\endcsname\relax\def\natexlab#1{#1}\fi
\expandafter\ifx\csname bibnamefont\endcsname\relax
  \def\bibnamefont#1{#1}\fi
\expandafter\ifx\csname bibfnamefont\endcsname\relax
  \def\bibfnamefont#1{#1}\fi
\expandafter\ifx\csname citenamefont\endcsname\relax
  \def\citenamefont#1{#1}\fi
\expandafter\ifx\csname url\endcsname\relax
  \def\url#1{\texttt{#1}}\fi
\expandafter\ifx\csname urlprefix\endcsname\relax\def\urlprefix{URL }\fi
\providecommand{\bibinfo}[2]{#2}
\providecommand{\eprint}[2][]{\url{#2}}

\bibitem[{\citenamefont{Kitaev}(2006)}]{Kitaev2006}
\bibinfo{author}{\bibfnamefont{A.}~\bibnamefont{Kitaev}},
  \bibinfo{journal}{Ann. Phys. (NY)} \textbf{\bibinfo{volume}{321}},
  \bibinfo{pages}{2 } (\bibinfo{year}{2006}).

\bibitem[{\citenamefont{Jackeli and Khaliullin}(2009)}]{Jackeli2009}
\bibinfo{author}{\bibfnamefont{G.}~\bibnamefont{Jackeli}} \bibnamefont{and}
  \bibinfo{author}{\bibfnamefont{G.}~\bibnamefont{Khaliullin}},
  \bibinfo{journal}{Phys. Rev. Lett.} \textbf{\bibinfo{volume}{102}},
  \bibinfo{pages}{017205} (\bibinfo{year}{2009}).

\bibitem[{\citenamefont{Chaloupka et~al.}(2010)\citenamefont{Chaloupka,
  Jackeli, and Khaliullin}}]{Chaloupka2010}
\bibinfo{author}{\bibfnamefont{J.}~\bibnamefont{Chaloupka}},
  \bibinfo{author}{\bibfnamefont{G.}~\bibnamefont{Jackeli}}, \bibnamefont{and}
  \bibinfo{author}{\bibfnamefont{G.}~\bibnamefont{Khaliullin}},
  \bibinfo{journal}{Phys. Rev. Lett.} \textbf{\bibinfo{volume}{105}},
  \bibinfo{pages}{027204} (\bibinfo{year}{2010}).

\bibitem[{\citenamefont{Singh et~al.}(2012)\citenamefont{Singh, Manni, Reuther,
  Berlijn, Thomale, Ku, Trebst, and Gegenwart}}]{Singh2012}
\bibinfo{author}{\bibfnamefont{Y.}~\bibnamefont{Singh}},
  \bibinfo{author}{\bibfnamefont{S.}~\bibnamefont{Manni}},
  \bibinfo{author}{\bibfnamefont{J.}~\bibnamefont{Reuther}},
  \bibinfo{author}{\bibfnamefont{T.}~\bibnamefont{Berlijn}},
  \bibinfo{author}{\bibfnamefont{R.}~\bibnamefont{Thomale}},
  \bibinfo{author}{\bibfnamefont{W.}~\bibnamefont{Ku}},
  \bibinfo{author}{\bibfnamefont{S.}~\bibnamefont{Trebst}}, \bibnamefont{and}
  \bibinfo{author}{\bibfnamefont{P.}~\bibnamefont{Gegenwart}},
  \bibinfo{journal}{Phys. Rev. Lett.} \textbf{\bibinfo{volume}{108}},
  \bibinfo{pages}{127203} (\bibinfo{year}{2012}).

\bibitem[{\citenamefont{Plumb et~al.}(2014)\citenamefont{Plumb, Clancy,
  Sandilands, Shankar, Hu, Burch, Kee, and Kim}}]{Plumb2014}
\bibinfo{author}{\bibfnamefont{K.~W.} \bibnamefont{Plumb}},
  \bibinfo{author}{\bibfnamefont{J.~P.} \bibnamefont{Clancy}},
  \bibinfo{author}{\bibfnamefont{L.~J.} \bibnamefont{Sandilands}},
  \bibinfo{author}{\bibfnamefont{V.~V.} \bibnamefont{Shankar}},
  \bibinfo{author}{\bibfnamefont{Y.~F.} \bibnamefont{Hu}},
  \bibinfo{author}{\bibfnamefont{K.~S.} \bibnamefont{Burch}},
  \bibinfo{author}{\bibfnamefont{H.-Y.} \bibnamefont{Kee}}, \bibnamefont{and}
  \bibinfo{author}{\bibfnamefont{Y.-J.} \bibnamefont{Kim}},
  \bibinfo{journal}{Phys. Rev. B} \textbf{\bibinfo{volume}{90}},
  \bibinfo{pages}{041112} (\bibinfo{year}{2014}).

\bibitem[{\citenamefont{Singh and Gegenwart}(2010)}]{Singh2010}
\bibinfo{author}{\bibfnamefont{Y.}~\bibnamefont{Singh}} \bibnamefont{and}
  \bibinfo{author}{\bibfnamefont{P.}~\bibnamefont{Gegenwart}},
  \bibinfo{journal}{Phys. Rev. B} \textbf{\bibinfo{volume}{82}},
  \bibinfo{pages}{064412} (\bibinfo{year}{2010}).

\bibitem[{\citenamefont{Sears et~al.}(2015)\citenamefont{Sears, Songvilay,
  Plumb, Clancy, Qiu, Zhao, Parshall, and Kim}}]{Sears2015}
\bibinfo{author}{\bibfnamefont{J.~A.} \bibnamefont{Sears}},
  \bibinfo{author}{\bibfnamefont{M.}~\bibnamefont{Songvilay}},
  \bibinfo{author}{\bibfnamefont{K.~W.} \bibnamefont{Plumb}},
  \bibinfo{author}{\bibfnamefont{J.~P.} \bibnamefont{Clancy}},
  \bibinfo{author}{\bibfnamefont{Y.}~\bibnamefont{Qiu}},
  \bibinfo{author}{\bibfnamefont{Y.}~\bibnamefont{Zhao}},
  \bibinfo{author}{\bibfnamefont{D.}~\bibnamefont{Parshall}}, \bibnamefont{and}
  \bibinfo{author}{\bibfnamefont{Y.-J.} \bibnamefont{Kim}},
  \bibinfo{journal}{Phys. Rev. B} \textbf{\bibinfo{volume}{91}},
  \bibinfo{pages}{144420} (\bibinfo{year}{2015}).

\bibitem[{\citenamefont{Williams et~al.}(2016)\citenamefont{Williams, Johnson,
  Freund, Choi, Jesche, Kimchi, Manni, Bombardi, Manuel, Gegenwart
  et~al.}}]{Williams2016}
\bibinfo{author}{\bibfnamefont{S.~C.} \bibnamefont{Williams}},
  \bibinfo{author}{\bibfnamefont{R.~D.} \bibnamefont{Johnson}},
  \bibinfo{author}{\bibfnamefont{F.}~\bibnamefont{Freund}},
  \bibinfo{author}{\bibfnamefont{S.}~\bibnamefont{Choi}},
  \bibinfo{author}{\bibfnamefont{A.}~\bibnamefont{Jesche}},
  \bibinfo{author}{\bibfnamefont{I.}~\bibnamefont{Kimchi}},
  \bibinfo{author}{\bibfnamefont{S.}~\bibnamefont{Manni}},
  \bibinfo{author}{\bibfnamefont{A.}~\bibnamefont{Bombardi}},
  \bibinfo{author}{\bibfnamefont{P.}~\bibnamefont{Manuel}},
  \bibinfo{author}{\bibfnamefont{P.}~\bibnamefont{Gegenwart}},
  \bibnamefont{et~al.}, \bibinfo{journal}{Phys. Rev. B}
  \textbf{\bibinfo{volume}{93}}, \bibinfo{pages}{195158}
  (\bibinfo{year}{2016}).

\bibitem[{\citenamefont{Sizyuk et~al.}(2014)\citenamefont{Sizyuk, Price,
  W\"olfle, and Perkins}}]{Sizyuk2014}
\bibinfo{author}{\bibfnamefont{Y.}~\bibnamefont{Sizyuk}},
  \bibinfo{author}{\bibfnamefont{C.}~\bibnamefont{Price}},
  \bibinfo{author}{\bibfnamefont{P.}~\bibnamefont{W\"olfle}}, \bibnamefont{and}
  \bibinfo{author}{\bibfnamefont{N.~B.} \bibnamefont{Perkins}},
  \bibinfo{journal}{Phys. Rev. B} \textbf{\bibinfo{volume}{90}},
  \bibinfo{pages}{155126} (\bibinfo{year}{2014}).

\bibitem[{\citenamefont{Kimchi and You}(2011)}]{Kimchi2011}
\bibinfo{author}{\bibfnamefont{I.}~\bibnamefont{Kimchi}} \bibnamefont{and}
  \bibinfo{author}{\bibfnamefont{Y.-Z.} \bibnamefont{You}},
  \bibinfo{journal}{Phys. Rev. B} \textbf{\bibinfo{volume}{84}},
  \bibinfo{pages}{180407} (\bibinfo{year}{2011}).

\bibitem[{\citenamefont{Rau et~al.}(2014)\citenamefont{Rau, Lee, and
  Kee}}]{Rau2014}
\bibinfo{author}{\bibfnamefont{J.~G.} \bibnamefont{Rau}},
  \bibinfo{author}{\bibfnamefont{E.~K.-H.} \bibnamefont{Lee}},
  \bibnamefont{and} \bibinfo{author}{\bibfnamefont{H.-Y.} \bibnamefont{Kee}},
  \bibinfo{journal}{Phys. Rev. Lett.} \textbf{\bibinfo{volume}{112}},
  \bibinfo{pages}{077204} (\bibinfo{year}{2014}).

\bibitem[{\citenamefont{Chaloupka and Khaliullin}(2015)}]{Chaloupka2015}
\bibinfo{author}{\bibfnamefont{J.}~\bibnamefont{Chaloupka}} \bibnamefont{and}
  \bibinfo{author}{\bibfnamefont{G.}~\bibnamefont{Khaliullin}},
  \bibinfo{journal}{Phys. Rev. B} \textbf{\bibinfo{volume}{92}},
  \bibinfo{pages}{024413} (\bibinfo{year}{2015}).

\bibitem[{\citenamefont{Sizyuk et~al.}(2016)\citenamefont{Sizyuk, W\"olfle, and
  Perkins}}]{Sizyuk2016}
\bibinfo{author}{\bibfnamefont{Y.}~\bibnamefont{Sizyuk}},
  \bibinfo{author}{\bibfnamefont{P.}~\bibnamefont{W\"olfle}}, \bibnamefont{and}
  \bibinfo{author}{\bibfnamefont{N.~B.} \bibnamefont{Perkins}},
  \bibinfo{journal}{Phys. Rev. B} \textbf{\bibinfo{volume}{94}},
  \bibinfo{pages}{085109} (\bibinfo{year}{2016}).

\bibitem[{\citenamefont{Winter et~al.}(2016)\citenamefont{Winter, Li, Jeschke,
  and Valent\'{\i}}}]{Winter2016}
\bibinfo{author}{\bibfnamefont{S.~M.} \bibnamefont{Winter}},
  \bibinfo{author}{\bibfnamefont{Y.}~\bibnamefont{Li}},
  \bibinfo{author}{\bibfnamefont{H.~O.} \bibnamefont{Jeschke}},
  \bibnamefont{and}
  \bibinfo{author}{\bibfnamefont{R.}~\bibnamefont{Valent\'{\i}}},
  \bibinfo{journal}{Phys. Rev. B} \textbf{\bibinfo{volume}{93}},
  \bibinfo{pages}{214431} (\bibinfo{year}{2016}).

\bibitem[{\citenamefont{Banerjee et~al.}(2016)\citenamefont{Banerjee, Bridges,
  Aczel, Li, Stone, Granroth, Lumsden, Yiu, Knolle, Bhattacharjee
  et~al.}}]{Banerjee2016}
\bibinfo{author}{\bibfnamefont{A.}~\bibnamefont{Banerjee}},
  \bibinfo{author}{\bibfnamefont{J.-Q.} \bibnamefont{Bridges},
  \bibfnamefont{C.~A.and~Yan}}, \bibinfo{author}{\bibfnamefont{A.~A.}
  \bibnamefont{Aczel}}, \bibinfo{author}{\bibfnamefont{L.}~\bibnamefont{Li}},
  \bibinfo{author}{\bibfnamefont{M.~B.} \bibnamefont{Stone}},
  \bibinfo{author}{\bibfnamefont{G.~E.} \bibnamefont{Granroth}},
  \bibinfo{author}{\bibfnamefont{M.~D.} \bibnamefont{Lumsden}},
  \bibinfo{author}{\bibfnamefont{Y.}~\bibnamefont{Yiu}},
  \bibinfo{author}{\bibfnamefont{J.}~\bibnamefont{Knolle}},
  \bibinfo{author}{\bibfnamefont{S.}~\bibnamefont{Bhattacharjee}},
  \bibnamefont{et~al.}, \bibinfo{journal}{Nat. Mater.}  (\bibinfo{year}{2016}).

\bibitem[{\citenamefont{Nasu et~al.}(2016)\citenamefont{Nasu, Knolle,
  Kovrizhin, Motome, and Moessner}}]{Nasu2016}
\bibinfo{author}{\bibfnamefont{J.}~\bibnamefont{Nasu}},
  \bibinfo{author}{\bibfnamefont{J.}~\bibnamefont{Knolle}},
  \bibinfo{author}{\bibfnamefont{D.~L.} \bibnamefont{Kovrizhin}},
  \bibinfo{author}{\bibfnamefont{Y.}~\bibnamefont{Motome}}, \bibnamefont{and}
  \bibinfo{author}{\bibfnamefont{R.}~\bibnamefont{Moessner}},
  \bibinfo{journal}{Nat. Phys.} \textbf{\bibinfo{volume}{12}},
  \bibinfo{pages}{912} (\bibinfo{year}{2016}), \bibinfo{note}{letter}.

\bibitem[{\citenamefont{{Trebst}}(2017)}]{Trebst2017}
\bibinfo{author}{\bibfnamefont{S.}~\bibnamefont{{Trebst}}},
  \bibinfo{journal}{ArXiv e-prints}  (\bibinfo{year}{2017}),
  \eprint{1701.07056}.

\bibitem[{\citenamefont{Baek et~al.}(2017)\citenamefont{Baek, Do, Choi, Kwon,
  Wolter, Nishimoto, van~den Brink, and B\"uchner}}]{Baek2017}
\bibinfo{author}{\bibfnamefont{S.-H.} \bibnamefont{Baek}},
  \bibinfo{author}{\bibfnamefont{S.-H.} \bibnamefont{Do}},
  \bibinfo{author}{\bibfnamefont{K.-Y.} \bibnamefont{Choi}},
  \bibinfo{author}{\bibfnamefont{Y.~S.} \bibnamefont{Kwon}},
  \bibinfo{author}{\bibfnamefont{A.~U.~B.} \bibnamefont{Wolter}},
  \bibinfo{author}{\bibfnamefont{S.}~\bibnamefont{Nishimoto}},
  \bibinfo{author}{\bibfnamefont{J.}~\bibnamefont{van~den Brink}},
  \bibnamefont{and}
  \bibinfo{author}{\bibfnamefont{B.}~\bibnamefont{B\"uchner}},
  \bibinfo{journal}{Phys. Rev. Lett.} \textbf{\bibinfo{volume}{119}},
  \bibinfo{pages}{037201} (\bibinfo{year}{2017}).

\bibitem[{\citenamefont{Leahy et~al.}(2017)\citenamefont{Leahy, Pocs,
  Siegfried, Graf, Do, Choi, Normand, and Lee}}]{Leahy2017}
\bibinfo{author}{\bibfnamefont{I.~A.} \bibnamefont{Leahy}},
  \bibinfo{author}{\bibfnamefont{C.~A.} \bibnamefont{Pocs}},
  \bibinfo{author}{\bibfnamefont{P.~E.} \bibnamefont{Siegfried}},
  \bibinfo{author}{\bibfnamefont{D.}~\bibnamefont{Graf}},
  \bibinfo{author}{\bibfnamefont{S.-H.} \bibnamefont{Do}},
  \bibinfo{author}{\bibfnamefont{K.-Y.} \bibnamefont{Choi}},
  \bibinfo{author}{\bibfnamefont{B.}~\bibnamefont{Normand}}, \bibnamefont{and}
  \bibinfo{author}{\bibfnamefont{M.}~\bibnamefont{Lee}},
  \bibinfo{journal}{Phys. Rev. Lett.} \textbf{\bibinfo{volume}{118}},
  \bibinfo{pages}{187203} (\bibinfo{year}{2017}).

\bibitem[{\citenamefont{Lee et~al.}(2006)\citenamefont{Lee, Nagaosa, and
  Wen}}]{Lee2006}
\bibinfo{author}{\bibfnamefont{P.~A.} \bibnamefont{Lee}},
  \bibinfo{author}{\bibfnamefont{N.}~\bibnamefont{Nagaosa}}, \bibnamefont{and}
  \bibinfo{author}{\bibfnamefont{X.-G.} \bibnamefont{Wen}},
  \bibinfo{journal}{Rev. Mod. Phys.} \textbf{\bibinfo{volume}{78}},
  \bibinfo{pages}{17} (\bibinfo{year}{2006}).

\bibitem[{\citenamefont{Edegger et~al.}(2007)\citenamefont{Edegger, Muthukumar,
  and Gros}}]{Edegger2007}
\bibinfo{author}{\bibfnamefont{B.}~\bibnamefont{Edegger}},
  \bibinfo{author}{\bibfnamefont{V.~N.} \bibnamefont{Muthukumar}},
  \bibnamefont{and} \bibinfo{author}{\bibfnamefont{C.}~\bibnamefont{Gros}},
  \bibinfo{journal}{Adv. Phys.} \textbf{\bibinfo{volume}{56}},
  \bibinfo{pages}{927} (\bibinfo{year}{2007}).

\bibitem[{\citenamefont{Ogata and Fukuyama}(2008)}]{Ogata2008}
\bibinfo{author}{\bibfnamefont{M.}~\bibnamefont{Ogata}} \bibnamefont{and}
  \bibinfo{author}{\bibfnamefont{H.}~\bibnamefont{Fukuyama}},
  \bibinfo{journal}{Rep. Prog. Phys.} \textbf{\bibinfo{volume}{71}},
  \bibinfo{pages}{036501} (\bibinfo{year}{2008}).

\bibitem[{\citenamefont{Hur and Rice}(2009)}]{LeHur2009}
\bibinfo{author}{\bibfnamefont{K.~L.} \bibnamefont{Hur}} \bibnamefont{and}
  \bibinfo{author}{\bibfnamefont{T.~M.} \bibnamefont{Rice}},
  \bibinfo{journal}{Ann. Phys. (NY)} \textbf{\bibinfo{volume}{324}},
  \bibinfo{pages}{1452 } (\bibinfo{year}{2009}).

\bibitem[{\citenamefont{Hyart et~al.}(2012)\citenamefont{Hyart, Wright,
  Khaliullin, and Rosenow}}]{Hyart2012}
\bibinfo{author}{\bibfnamefont{T.}~\bibnamefont{Hyart}},
  \bibinfo{author}{\bibfnamefont{A.~R.} \bibnamefont{Wright}},
  \bibinfo{author}{\bibfnamefont{G.}~\bibnamefont{Khaliullin}},
  \bibnamefont{and} \bibinfo{author}{\bibfnamefont{B.}~\bibnamefont{Rosenow}},
  \bibinfo{journal}{Phys. Rev. B} \textbf{\bibinfo{volume}{85}},
  \bibinfo{pages}{140510} (\bibinfo{year}{2012}).

\bibitem[{\citenamefont{You et~al.}(2012)\citenamefont{You, Kimchi, and
  Vishwanath}}]{You2012}
\bibinfo{author}{\bibfnamefont{Y.-Z.} \bibnamefont{You}},
  \bibinfo{author}{\bibfnamefont{I.}~\bibnamefont{Kimchi}}, \bibnamefont{and}
  \bibinfo{author}{\bibfnamefont{A.}~\bibnamefont{Vishwanath}},
  \bibinfo{journal}{Phys. Rev. B} \textbf{\bibinfo{volume}{86}},
  \bibinfo{pages}{085145} (\bibinfo{year}{2012}).

\bibitem[{\citenamefont{Okamoto}(2013)}]{Okamoto2013}
\bibinfo{author}{\bibfnamefont{S.}~\bibnamefont{Okamoto}},
  \bibinfo{journal}{Phys. Rev. B} \textbf{\bibinfo{volume}{87}},
  \bibinfo{pages}{064508} (\bibinfo{year}{2013}).

\bibitem[{\citenamefont{Scherer et~al.}(2014)\citenamefont{Scherer, Scherer,
  Khaliullin, Honerkamp, and Rosenow}}]{Scherer2014}
\bibinfo{author}{\bibfnamefont{D.~D.} \bibnamefont{Scherer}},
  \bibinfo{author}{\bibfnamefont{M.~M.} \bibnamefont{Scherer}},
  \bibinfo{author}{\bibfnamefont{G.}~\bibnamefont{Khaliullin}},
  \bibinfo{author}{\bibfnamefont{C.}~\bibnamefont{Honerkamp}},
  \bibnamefont{and} \bibinfo{author}{\bibfnamefont{B.}~\bibnamefont{Rosenow}},
  \bibinfo{journal}{Phys. Rev. B} \textbf{\bibinfo{volume}{90}},
  \bibinfo{pages}{045135} (\bibinfo{year}{2014}).

\bibitem[{\citenamefont{Hyart et~al.}(2014)\citenamefont{Hyart, Wright, and
  Rosenow}}]{Hyart2014}
\bibinfo{author}{\bibfnamefont{T.}~\bibnamefont{Hyart}},
  \bibinfo{author}{\bibfnamefont{A.~R.} \bibnamefont{Wright}},
  \bibnamefont{and} \bibinfo{author}{\bibfnamefont{B.}~\bibnamefont{Rosenow}},
  \bibinfo{journal}{Phys. Rev. B} \textbf{\bibinfo{volume}{90}},
  \bibinfo{pages}{064507} (\bibinfo{year}{2014}).

\bibitem[{\citenamefont{Kimme et~al.}(2015)\citenamefont{Kimme, Hyart, and
  Rosenow}}]{Kimme2015}
\bibinfo{author}{\bibfnamefont{L.}~\bibnamefont{Kimme}},
  \bibinfo{author}{\bibfnamefont{T.}~\bibnamefont{Hyart}}, \bibnamefont{and}
  \bibinfo{author}{\bibfnamefont{B.}~\bibnamefont{Rosenow}},
  \bibinfo{journal}{Phys. Rev. B} \textbf{\bibinfo{volume}{91}},
  \bibinfo{pages}{220501} (\bibinfo{year}{2015}).

\bibitem[{\citenamefont{Liu et~al.}(2016)\citenamefont{Liu, Repellin,
  Dou\ifmmode~\mbox{\c{c}}\else \c{c}\fi{}ot, Regnault, and Le~Hur}}]{Liu2016}
\bibinfo{author}{\bibfnamefont{T.}~\bibnamefont{Liu}},
  \bibinfo{author}{\bibfnamefont{C.}~\bibnamefont{Repellin}},
  \bibinfo{author}{\bibfnamefont{B.}~\bibnamefont{Dou\ifmmode~\mbox{\c{c}}\else
  \c{c}\fi{}ot}}, \bibinfo{author}{\bibfnamefont{N.}~\bibnamefont{Regnault}},
  \bibnamefont{and} \bibinfo{author}{\bibfnamefont{K.}~\bibnamefont{Le~Hur}},
  \bibinfo{journal}{Phys. Rev. B} \textbf{\bibinfo{volume}{94}},
  \bibinfo{pages}{180506} (\bibinfo{year}{2016}).

\bibitem[{\citenamefont{Baskaran}(2002)}]{Baskaran2002}
\bibinfo{author}{\bibfnamefont{G.}~\bibnamefont{Baskaran}},
  \bibinfo{journal}{Phys. Rev. B} \textbf{\bibinfo{volume}{65}},
  \bibinfo{pages}{212505} (\bibinfo{year}{2002}).

\bibitem[{\citenamefont{Black-Schaffer and Doniach}(2007)}]{BlackSchaffer2007}
\bibinfo{author}{\bibfnamefont{A.~M.} \bibnamefont{Black-Schaffer}}
  \bibnamefont{and} \bibinfo{author}{\bibfnamefont{S.}~\bibnamefont{Doniach}},
  \bibinfo{journal}{Phys. Rev. B} \textbf{\bibinfo{volume}{75}},
  \bibinfo{pages}{134512} (\bibinfo{year}{2007}).

\bibitem[{\citenamefont{Barnes}(1976)}]{Barnes1976}
\bibinfo{author}{\bibfnamefont{S.}~\bibnamefont{Barnes}}, \bibinfo{journal}{J.
  Phys. F} \textbf{\bibinfo{volume}{{6}}}, \bibinfo{pages}{1375}
  (\bibinfo{year}{1976}).

\bibitem[{\citenamefont{Baskaran et~al.}(1987)\citenamefont{Baskaran, Zou, and
  Anderson}}]{Baskaran1987}
\bibinfo{author}{\bibfnamefont{G.}~\bibnamefont{Baskaran}},
  \bibinfo{author}{\bibfnamefont{Z.}~\bibnamefont{Zou}}, \bibnamefont{and}
  \bibinfo{author}{\bibfnamefont{P.~W.} \bibnamefont{Anderson}},
  \bibinfo{journal}{Solid State Commun.} \textbf{\bibinfo{volume}{63}},
  \bibinfo{pages}{973} (\bibinfo{year}{1987}).

\bibitem[{\citenamefont{Honerkamp}(2008)}]{Honerkamp2008}
\bibinfo{author}{\bibfnamefont{C.}~\bibnamefont{Honerkamp}},
  \bibinfo{journal}{Phys. Rev. Lett.} \textbf{\bibinfo{volume}{100}},
  \bibinfo{pages}{146404} (\bibinfo{year}{2008}).

\bibitem[{\citenamefont{Wu et~al.}(2013)\citenamefont{Wu, Scherer, Honerkamp,
  and Le~Hur}}]{Wu2013}
\bibinfo{author}{\bibfnamefont{W.}~\bibnamefont{Wu}},
  \bibinfo{author}{\bibfnamefont{M.~M.} \bibnamefont{Scherer}},
  \bibinfo{author}{\bibfnamefont{C.}~\bibnamefont{Honerkamp}},
  \bibnamefont{and} \bibinfo{author}{\bibfnamefont{K.}~\bibnamefont{Le~Hur}},
  \bibinfo{journal}{Phys. Rev. B} \textbf{\bibinfo{volume}{87}},
  \bibinfo{pages}{094521} (\bibinfo{year}{2013}).

\bibitem[{\citenamefont{Sigrist and Ueda}(1991)}]{Sigrist1991}
\bibinfo{author}{\bibfnamefont{M.}~\bibnamefont{Sigrist}} \bibnamefont{and}
  \bibinfo{author}{\bibfnamefont{K.}~\bibnamefont{Ueda}},
  \bibinfo{journal}{Rev. Mod. Phys.} \textbf{\bibinfo{volume}{63}},
  \bibinfo{pages}{239} (\bibinfo{year}{1991}).

\bibitem[{\citenamefont{Yamaji et~al.}(2014)\citenamefont{Yamaji, Nomura,
  Kurita, Arita, and Imada}}]{Yamaji2014}
\bibinfo{author}{\bibfnamefont{Y.}~\bibnamefont{Yamaji}},
  \bibinfo{author}{\bibfnamefont{Y.}~\bibnamefont{Nomura}},
  \bibinfo{author}{\bibfnamefont{M.}~\bibnamefont{Kurita}},
  \bibinfo{author}{\bibfnamefont{R.}~\bibnamefont{Arita}}, \bibnamefont{and}
  \bibinfo{author}{\bibfnamefont{M.}~\bibnamefont{Imada}},
  \bibinfo{journal}{Phys. Rev. Lett.} \textbf{\bibinfo{volume}{113}},
  \bibinfo{pages}{107201} (\bibinfo{year}{2014}).

\bibitem[{\citenamefont{Hu et~al.}(2015)\citenamefont{Hu, Wang, and
  Feng}}]{Hu2015}
\bibinfo{author}{\bibfnamefont{K.}~\bibnamefont{Hu}},
  \bibinfo{author}{\bibfnamefont{F.}~\bibnamefont{Wang}}, \bibnamefont{and}
  \bibinfo{author}{\bibfnamefont{J.}~\bibnamefont{Feng}},
  \bibinfo{journal}{Phys. Rev. Lett.} \textbf{\bibinfo{volume}{115}},
  \bibinfo{pages}{167204} (\bibinfo{year}{2015}).

\bibitem[{\citenamefont{Kim et~al.}(2015)\citenamefont{Kim, V., Catuneanu, and
  Kee}}]{Kim2015-RuCl}
\bibinfo{author}{\bibfnamefont{H.-S.} \bibnamefont{Kim}},
  \bibinfo{author}{\bibfnamefont{V.~S.} \bibnamefont{V.}},
  \bibinfo{author}{\bibfnamefont{A.}~\bibnamefont{Catuneanu}},
  \bibnamefont{and} \bibinfo{author}{\bibfnamefont{H.-Y.} \bibnamefont{Kee}},
  \bibinfo{journal}{Phys. Rev. B} \textbf{\bibinfo{volume}{91}},
  \bibinfo{pages}{241110} (\bibinfo{year}{2015}).

\bibitem[{\citenamefont{Yamaji et~al.}(2016)\citenamefont{Yamaji, Suzuki,
  Yamada, Suga, Kawashima, and Imada}}]{Yamaji2016}
\bibinfo{author}{\bibfnamefont{Y.}~\bibnamefont{Yamaji}},
  \bibinfo{author}{\bibfnamefont{T.}~\bibnamefont{Suzuki}},
  \bibinfo{author}{\bibfnamefont{T.}~\bibnamefont{Yamada}},
  \bibinfo{author}{\bibfnamefont{S.-i.} \bibnamefont{Suga}},
  \bibinfo{author}{\bibfnamefont{N.}~\bibnamefont{Kawashima}},
  \bibnamefont{and} \bibinfo{author}{\bibfnamefont{M.}~\bibnamefont{Imada}},
  \bibinfo{journal}{Phys. Rev. B} \textbf{\bibinfo{volume}{93}},
  \bibinfo{pages}{174425} (\bibinfo{year}{2016}).

\bibitem[{\citenamefont{Nishimoto et~al.}(2016)\citenamefont{Nishimoto,
  Katukuri, Yushankhai, Stoll, Roszler, Hozoi, Rousochatzakis, and van~den
  Brink}}]{Nishimoto2016}
\bibinfo{author}{\bibfnamefont{S.}~\bibnamefont{Nishimoto}},
  \bibinfo{author}{\bibfnamefont{V.~M.} \bibnamefont{Katukuri}},
  \bibinfo{author}{\bibfnamefont{V.}~\bibnamefont{Yushankhai}},
  \bibinfo{author}{\bibfnamefont{H.}~\bibnamefont{Stoll}},
  \bibinfo{author}{\bibfnamefont{U.~K.} \bibnamefont{Roszler}},
  \bibinfo{author}{\bibfnamefont{L.}~\bibnamefont{Hozoi}},
  \bibinfo{author}{\bibfnamefont{I.}~\bibnamefont{Rousochatzakis}},
  \bibnamefont{and} \bibinfo{author}{\bibfnamefont{J.}~\bibnamefont{van~den
  Brink}}, \bibinfo{journal}{Nat. Commun.} \textbf{\bibinfo{volume}{7}}
  (\bibinfo{year}{2016}).

\bibitem[{\citenamefont{Ran et~al.}(2017)\citenamefont{Ran, Wang, Wang, Dong,
  Ren, Bao, Li, Ma, Gan, Zhang et~al.}}]{Ran2017}
\bibinfo{author}{\bibfnamefont{K.}~\bibnamefont{Ran}},
  \bibinfo{author}{\bibfnamefont{J.}~\bibnamefont{Wang}},
  \bibinfo{author}{\bibfnamefont{W.}~\bibnamefont{Wang}},
  \bibinfo{author}{\bibfnamefont{Z.-Y.} \bibnamefont{Dong}},
  \bibinfo{author}{\bibfnamefont{X.}~\bibnamefont{Ren}},
  \bibinfo{author}{\bibfnamefont{S.}~\bibnamefont{Bao}},
  \bibinfo{author}{\bibfnamefont{S.}~\bibnamefont{Li}},
  \bibinfo{author}{\bibfnamefont{Z.}~\bibnamefont{Ma}},
  \bibinfo{author}{\bibfnamefont{Y.}~\bibnamefont{Gan}},
  \bibinfo{author}{\bibfnamefont{Y.}~\bibnamefont{Zhang}},
  \bibnamefont{et~al.}, \bibinfo{journal}{Phys. Rev. Lett.}
  \textbf{\bibinfo{volume}{118}}, \bibinfo{pages}{107203}
  (\bibinfo{year}{2017}).

\bibitem[{\citenamefont{Janssen et~al.}(2017)\citenamefont{Janssen, Andrade,
  and Vojta}}]{Janssen2017}
\bibinfo{author}{\bibfnamefont{L.}~\bibnamefont{Janssen}},
  \bibinfo{author}{\bibfnamefont{E.~C.} \bibnamefont{Andrade}},
  \bibnamefont{and} \bibinfo{author}{\bibfnamefont{M.}~\bibnamefont{Vojta}},
  \bibinfo{journal}{Phys. Rev. B} \textbf{\bibinfo{volume}{96}},
  \bibinfo{pages}{064430} (\bibinfo{year}{2017}).

\bibitem[{\citenamefont{Katukuri et~al.}(2014)\citenamefont{Katukuri,
  Nishimoto, Yushankhai, Stoyanova, Kandpal, Choi, Coldea, Rousochatzakis,
  Hozoi, and van~den Brink}}]{Katukuri2014}
\bibinfo{author}{\bibfnamefont{V.~M.} \bibnamefont{Katukuri}},
  \bibinfo{author}{\bibfnamefont{S.}~\bibnamefont{Nishimoto}},
  \bibinfo{author}{\bibfnamefont{V.}~\bibnamefont{Yushankhai}},
  \bibinfo{author}{\bibfnamefont{A.}~\bibnamefont{Stoyanova}},
  \bibinfo{author}{\bibfnamefont{H.}~\bibnamefont{Kandpal}},
  \bibinfo{author}{\bibfnamefont{S.}~\bibnamefont{Choi}},
  \bibinfo{author}{\bibfnamefont{R.}~\bibnamefont{Coldea}},
  \bibinfo{author}{\bibfnamefont{I.}~\bibnamefont{Rousochatzakis}},
  \bibinfo{author}{\bibfnamefont{L.}~\bibnamefont{Hozoi}}, \bibnamefont{and}
  \bibinfo{author}{\bibfnamefont{J.}~\bibnamefont{van~den Brink}},
  \bibinfo{journal}{New J. Phys.} \textbf{\bibinfo{volume}{16}},
  \bibinfo{pages}{013056} (\bibinfo{year}{2014}).

\bibitem[{\citenamefont{Yadav et~al.}(2016)\citenamefont{Yadav, Bogdanov,
  Katukuri, Nishimoto, van~den Brink, and Hozoi}}]{Yadav2016}
\bibinfo{author}{\bibfnamefont{R.}~\bibnamefont{Yadav}},
  \bibinfo{author}{\bibfnamefont{N.~A.} \bibnamefont{Bogdanov}},
  \bibinfo{author}{\bibfnamefont{V.~M.} \bibnamefont{Katukuri}},
  \bibinfo{author}{\bibfnamefont{S.}~\bibnamefont{Nishimoto}},
  \bibinfo{author}{\bibfnamefont{J.}~\bibnamefont{van~den Brink}},
  \bibnamefont{and} \bibinfo{author}{\bibfnamefont{L.}~\bibnamefont{Hozoi}},
  \bibinfo{journal}{Sci. Rep.} \textbf{\bibinfo{volume}{6}}
  (\bibinfo{year}{2016}).

\bibitem[{\citenamefont{Kim and Kee}(2016)}]{Kim2016}
\bibinfo{author}{\bibfnamefont{H.-S.} \bibnamefont{Kim}} \bibnamefont{and}
  \bibinfo{author}{\bibfnamefont{H.-Y.} \bibnamefont{Kee}},
  \bibinfo{journal}{Phys. Rev. B} \textbf{\bibinfo{volume}{93}},
  \bibinfo{pages}{155143} (\bibinfo{year}{2016}).

\bibitem[{\citenamefont{Schnyder et~al.}(2008)\citenamefont{Schnyder, Ryu,
  Furusaki, and Ludwig}}]{Schnyder2008}
\bibinfo{author}{\bibfnamefont{A.~P.} \bibnamefont{Schnyder}},
  \bibinfo{author}{\bibfnamefont{S.}~\bibnamefont{Ryu}},
  \bibinfo{author}{\bibfnamefont{A.}~\bibnamefont{Furusaki}}, \bibnamefont{and}
  \bibinfo{author}{\bibfnamefont{A.~W.~W.} \bibnamefont{Ludwig}},
  \bibinfo{journal}{Phys. Rev. B} \textbf{\bibinfo{volume}{78}},
  \bibinfo{pages}{195125} (\bibinfo{year}{2008}).

\bibitem[{\citenamefont{Fukui et~al.}(2005)\citenamefont{Fukui, Hatsugai, and
  Suzuki}}]{Fukui2005}
\bibinfo{author}{\bibfnamefont{T.}~\bibnamefont{Fukui}},
  \bibinfo{author}{\bibfnamefont{Y.}~\bibnamefont{Hatsugai}}, \bibnamefont{and}
  \bibinfo{author}{\bibfnamefont{H.}~\bibnamefont{Suzuki}},
  \bibinfo{journal}{J. Phys. Soc. Jpn.} \textbf{\bibinfo{volume}{74}},
  \bibinfo{pages}{1674} (\bibinfo{year}{2005}).

\bibitem[{\citenamefont{L\"othman and Black-Schaffer}(2014)}]{Lothman2014}
\bibinfo{author}{\bibfnamefont{T.}~\bibnamefont{L\"othman}} \bibnamefont{and}
  \bibinfo{author}{\bibfnamefont{A.~M.} \bibnamefont{Black-Schaffer}},
  \bibinfo{journal}{Phys. Rev. B} \textbf{\bibinfo{volume}{90}},
  \bibinfo{pages}{224504} (\bibinfo{year}{2014}).

\bibitem[{\citenamefont{Sato}(2009)}]{Sato2009}
\bibinfo{author}{\bibfnamefont{M.}~\bibnamefont{Sato}}, \bibinfo{journal}{Phys.
  Rev. B} \textbf{\bibinfo{volume}{79}}, \bibinfo{pages}{214526}
  (\bibinfo{year}{2009}).

\bibitem[{\citenamefont{Sato}(2010)}]{Sato2010}
\bibinfo{author}{\bibfnamefont{M.}~\bibnamefont{Sato}}, \bibinfo{journal}{Phys.
  Rev. B} \textbf{\bibinfo{volume}{81}}, \bibinfo{pages}{220504}
  (\bibinfo{year}{2010}).

\bibitem[{Note1()}]{Note1}
Note1, \bibinfo{note}{numerically we find very near degenerate states, as a
  perfect energy degeneracy is hard to achieve due to finite sampling in the
  Brillouin zone.}

\bibitem[{\citenamefont{Venderbos et~al.}(2016)\citenamefont{Venderbos, Kozii,
  and Fu}}]{Venderbos2016}
\bibinfo{author}{\bibfnamefont{J.~W.~F.} \bibnamefont{Venderbos}},
  \bibinfo{author}{\bibfnamefont{V.}~\bibnamefont{Kozii}}, \bibnamefont{and}
  \bibinfo{author}{\bibfnamefont{L.}~\bibnamefont{Fu}}, \bibinfo{journal}{Phys.
  Rev. B} \textbf{\bibinfo{volume}{94}}, \bibinfo{pages}{180504}
  (\bibinfo{year}{2016}).

\bibitem[{\citenamefont{Shitade et~al.}(2009)\citenamefont{Shitade, Katsura,
  Kune\ifmmode~\check{s}\else \v{s}\fi{}, Qi, Zhang, and
  Nagaosa}}]{Shitade2009}
\bibinfo{author}{\bibfnamefont{A.}~\bibnamefont{Shitade}},
  \bibinfo{author}{\bibfnamefont{H.}~\bibnamefont{Katsura}},
  \bibinfo{author}{\bibfnamefont{J.}~\bibnamefont{Kune\ifmmode~\check{s}\else
  \v{s}\fi{}}}, \bibinfo{author}{\bibfnamefont{X.-L.} \bibnamefont{Qi}},
  \bibinfo{author}{\bibfnamefont{S.-C.} \bibnamefont{Zhang}}, \bibnamefont{and}
  \bibinfo{author}{\bibfnamefont{N.}~\bibnamefont{Nagaosa}},
  \bibinfo{journal}{Phys. Rev. Lett.} \textbf{\bibinfo{volume}{102}},
  \bibinfo{pages}{256403} (\bibinfo{year}{2009}).

\bibitem[{\citenamefont{Mazin et~al.}(2012)\citenamefont{Mazin, Jeschke,
  Foyevtsova, Valent\'\i, and Khomskii}}]{Mazin2012}
\bibinfo{author}{\bibfnamefont{I.~I.} \bibnamefont{Mazin}},
  \bibinfo{author}{\bibfnamefont{H.~O.} \bibnamefont{Jeschke}},
  \bibinfo{author}{\bibfnamefont{K.}~\bibnamefont{Foyevtsova}},
  \bibinfo{author}{\bibfnamefont{R.}~\bibnamefont{Valent\'\i}},
  \bibnamefont{and} \bibinfo{author}{\bibfnamefont{D.~I.}
  \bibnamefont{Khomskii}}, \bibinfo{journal}{Phys. Rev. Lett.}
  \textbf{\bibinfo{volume}{109}}, \bibinfo{pages}{197201}
  (\bibinfo{year}{2012}).

\bibitem[{\citenamefont{Foyevtsova et~al.}(2013)\citenamefont{Foyevtsova,
  Jeschke, Mazin, Khomskii, and Valent\'\i}}]{Foyevtsova2013}
\bibinfo{author}{\bibfnamefont{K.}~\bibnamefont{Foyevtsova}},
  \bibinfo{author}{\bibfnamefont{H.~O.} \bibnamefont{Jeschke}},
  \bibinfo{author}{\bibfnamefont{I.~I.} \bibnamefont{Mazin}},
  \bibinfo{author}{\bibfnamefont{D.~I.} \bibnamefont{Khomskii}},
  \bibnamefont{and}
  \bibinfo{author}{\bibfnamefont{R.}~\bibnamefont{Valent\'\i}},
  \bibinfo{journal}{Phys. Rev. B} \textbf{\bibinfo{volume}{88}},
  \bibinfo{pages}{035107} (\bibinfo{year}{2013}).

\end{thebibliography}

\end{document}